\documentclass[
reprint,
superscriptaddress,
amsmath,amssymb,
aps,
pra,
floatfix,
]{revtex4-2}

\usepackage{hyperref}  \hypersetup{colorlinks,allcolors=blue,unicode}

\usepackage[utf8]{inputenc}
\usepackage[T1]{fontenc}
\usepackage{microtype}
\usepackage{newtxtext,newtxmath}

\usepackage{graphicx}\usepackage{xcolor}
\usepackage{dcolumn}

\usepackage{mathtools}

\usepackage[retain-explicit-plus,list-final-separator={, and },range-phrase=\textendash,range-units=single]{siunitx}
\DeclareSIUnit\angstrom{\text{\AA}}

\usepackage[version=4]{mhchem}

\usepackage{booktabs}
\usepackage{multirow}
\newcolumntype{P}[1]{>{\centering\arraybackslash}p{#1}}

\usepackage{array}
\newcolumntype{H}{>{\setbox0=\hbox\bgroup}c<{\egroup}@{}}
\newcolumntype{C}[1]{>{\centering\arraybackslash}p{#1}}

\usepackage[capitalize,noabbrev,nameinlink]{cleveref}
\crefformat{equation}{#2eq~#1#3}  \crefmultiformat{equation}{\edef\crefstripprefixinfo{#1}#2eqs~#1#3}{ and~#2\crefstripprefix{\crefstripprefixinfo}{#1}#3}{, #2\crefstripprefix{\crefstripprefixinfo}{#1}#3}{, and~#2\crefstripprefix{\crefstripprefixinfo}{#1}#3}

\usepackage{enumitem}

\newcommand{\bfTBDMTP}[1][]{\textbf{TB+$\mathbf{\Delta}$MTP\##1}}
\newcommand{\ptitle}[1]{}

\usepackage[all]{hypcap}
\usepackage{float}
\usepackage{anyfontsize}
\usepackage{orcidlink}

\colorlet{rev0}{black}

\begin{document}  

\title{Machine-learning interatomic potentials achieving CCSD(T) accuracy for \textcolor{rev0}{systems with extended covalent networks and van der Waals interactions}}

\author{Yuji Ikeda\,\orcidlink{0000-0001-9176-3270}}
\email{yuji.ikeda@imw.uni-stuttgart.de}
\affiliation{Institute for Materials Science,
\href{https://ror.org/04vnq7t77}{University of Stuttgart},
Pfaffenwaldring 55, 70569 Stuttgart, Germany}

\author{Axel Forslund\,\orcidlink{0000-0002-0419-3546}}
\affiliation{Institute for Materials Science,
\href{https://ror.org/04vnq7t77}{University of Stuttgart},
Pfaffenwaldring 55, 70569 Stuttgart, Germany}
\affiliation{Department of Materials Science and Engineering,
\href{https://ror.org/026vcq606}{KTH Royal Institute of Technology},
SE-100 44 Stockholm, Sweden}

\author{Pranav Kumar\,\orcidlink{0000-0002-3661-5870}}
\affiliation{Institute for Materials Science,
\href{https://ror.org/04vnq7t77}{University of Stuttgart},
Pfaffenwaldring 55, 70569 Stuttgart, Germany}

\author{Yongliang~Ou\,\orcidlink{0000-0003-1486-1197}}
\affiliation{Institute for Materials Science,
\href{https://ror.org/04vnq7t77}{University of Stuttgart},
Pfaffenwaldring 55, 70569 Stuttgart, Germany}
\affiliation{Department of Materials Science and Engineering,
\href{https://ror.org/042nb2s44}{Massachusetts Institute of Technology},
77 Massachusetts Avenue, Cambridge, 02139, MA, USA}

\author{Jong Hyun Jung\,\orcidlink{0000-0002-2409-975X}}
\affiliation{Institute for Materials Science,
\href{https://ror.org/04vnq7t77}{University of Stuttgart},
Pfaffenwaldring 55, 70569 Stuttgart, Germany}

\author{Andreas Köhn\,\orcidlink{0000-0002-0844-842X}}
\affiliation{Institute for Theoretical Chemistry,
\href{https://ror.org/04vnq7t77}{University of Stuttgart},
Pfaffenwaldring 55, 70569 Stuttgart, Germany}

\author{Blazej Grabowski\,\orcidlink{0000-0003-4281-5665}}
\affiliation{Institute for Materials Science,
\href{https://ror.org/04vnq7t77}{University of Stuttgart},
Pfaffenwaldring 55, 70569 Stuttgart, Germany}

\begin{abstract}

Machine-learning interatomic potentials (MLIPs) enable large-scale atomistic simulations at moderate computational cost while retaining \textit{ab initio} accuracy.
\color{rev0}
In recent years, MLIPs trained on coupled-cluster data---particularly CCSD(T), which includes single, double, and perturbative triple excitations---have emerged as a promising route to achieve chemical accuracy (\qty{1}{kcal/mol}) beyond the limits of density functional theory (DFT) and to incorporate non-empirical van der Waals (vdW) interactions.
Most existing approaches are, however, still not straightforwardly applicable for systems with extended covalent networks such as covalent organic frameworks (COFs) due to the limited availability of CCSD(T) under periodic boundary conditions.
Here we present a methodology to train MLIPs with CCSD(T) accuracy for systems with extended covalent networks.
The approach is based on the $\Delta$-learning method with a dispersion-corrected tight-binding baseline and an MLIP trained on the differences of the target CCSD(T) energies from the baseline.
This $\Delta$-learning strategy enables training on compact molecular fragments while preserving transferability toward the periodic systems.
Dispersion interactions are accounted for by including vdW-bound multimers in the training set, and the combination with a vdW-aware tight-binding baseline allows the formally local MLIP to attain CCSD(T)-level accuracy even for systems dominated by long-range vdW forces.
\color{black}
The resulting potential yields root‑mean‑square energy errors below \qty{0.4}{meV/atom} on both training and test sets and reproduces electronic total atomization energies, bond lengths, harmonic vibrational frequencies, and inter-molecular interaction energies for benchmark molecular systems.
We apply the method to a prototypical quasi-two-dimensional covalent organic framework (COF) composed of carbon and hydrogen.
The COF structure, inter-layer binding energies, and hydrogen absorption are analyzed at CCSD(T) accuracy.
\color{rev0}
Overall, the developed \textcolor{rev0}{methodology} opens a practical route to large‑scale atomistic simulations for systems with extended covalent networks and vdW interactions with chemical accuracy.
\color{black}

\end{abstract}

\maketitle  

\section{Introduction\label{sec:introduction}}

Machine-learning interatomic potentials (MLIPs)~\cite{Behler_JCP_2016_Perspective} are a new generation of interatomic potentials with flexible mathematics-oriented forms.
MLIPs can accurately reproduce complicated potential-energy surfaces (PESs) of \textit{ab initio} calculations, typically with errors below \qty{1}{meV/atom}~\cite{Grabowski_nCM_2019_Ab} and at a fraction of the computational cost.

Most MLIPs to date are trained on density-functional theory (DFT) data.
However, DFT suffers from errors due to the intrinsic approximations in exchange--correlation functionals and therefore often fails to reproduce experimental or higher-accuracy computational results~\cite{Erhard_PRL_2016_Power}.
Moreover, the local density approximation (LDA) and the generalized gradient approximation (GGA) themselves cannot capture long-range van der Waals (vdW) interactions even qualitatively due to their local or semi-local nature.
It is therefore common to add vdW interactions explicitly on top of these standard exchange--correlation functionals.
These vdW functionals can be divided into element-specific schemes (e.g., the D4 correction~\cite{Caldeweyher_JCP_2019_generally,  Ehlert_JCP_154_061101_2021})  and element-independent ones (e.g., the rVV10 correction~\cite{Sabatini_PRB_2013_Nonlocal,  Peng_PRX_2016_Versatile,  Ning_PRB_2022_Workhorse}).  However, both categories remain essentially semi-empirical, as their parameters are tuned for limited benchmark sets, which can restrict their transferability.

Wavefunction-based post-Hartree--Fock (HF) methods~\cite{Szabo_Book_1996_Modern,Helgaker_Book_2000_Molecular} can overcome the issues associated with DFT, because their accuracy can be improved systematically in a non-empirical manner and vdW interactions are intrinsically included.
Particularly, the coupled-cluster method with single, double, and perturbative triple excitations (CCSD(T))~\cite{Raghavachari_CPL_157_479_1989} can achieve the so-called chemical accuracy of \qty{1}{kcal/mol} ($\approx$\,\qty{40}{meV/system}), even below typical experimental errors, and is thus regarded as the gold standard of computational chemistry.
However, the computational cost of canonical CCSD(T) scales steeply as $\mathcal{O}(N^7)$ with the number of correlated orbitals $N$, restricting routine applications to systems containing only a few dozen atoms.

MLIPs offer a way to sidestep the drawbacks of both DFT and CCSD(T).
When trained on CCSD(T) reference data, MLIPs inherit its chemical accuracy and naturally incorporate vdW interactions while retaining the near-linear scaling of classical force fields.
\textcolor{rev0}{Various} CCSD(T)-\textcolor{rev0}{level} MLIPs have already been reported~~\cite{
Schran_JCP_148_102310_2017,  Chmiela_NC_9_3887_2018,
Schran_JCTC_14_5068_2018,
Sauceda_JCP_150_114102_2019,
Schran_JCTC_16_88_2019,
Smith_NC_2019_Approaching,Smith_SD_2020_ANI,
Bogojeski_NC_11_5223_2020,
Topolnicki_JCTC_16_6785_2020,
Schran_JCP_154_51101_2021,
Beckmann_JCTC_18_5492_2022,  Daru_PRL_129_226001_2022,
Yu_JPCL_13_5068_2022,  Qu_JCTC_19_3446_2023,  Zhu_JCTC_19_3551_2023,  Chen_JCTC_19_4510_2023,
Herzog_nCM_10_68_2024,
Kim_JACS_147_1042_2024,
Tang_NCS_5_144_2024,*Tang_NCS_5_184_2025,
ONeill____2025}.
One popular line of research trains MLIPs on CCSD(T) potentials for small molecules or molecular clusters such as protonated water clusters~\cite{
Schran_JCP_148_102310_2017,  Schran_JCTC_14_5068_2018,
Schran_JCTC_16_88_2019,
Topolnicki_JCTC_16_6785_2020,
Schran_JCP_154_51101_2021,
Beckmann_JCTC_18_5492_2022,  Chmiela_NC_9_3887_2018,
Sauceda_JCP_150_114102_2019};
another targets liquid water~\cite{
Yu_JPCL_13_5068_2022,  Qu_JCTC_19_3446_2023,  Zhu_JCTC_19_3551_2023,  Daru_PRL_129_226001_2022,
Chen_JCTC_19_4510_2023,
ONeill____2025}.
For wider chemical coverage, Smith~\textit{et al.}~\cite{Smith_NC_2019_Approaching,Smith_SD_2020_ANI} developed the ANI-1ccx potential trained on \num{500000} configurations of monomer molecules computed at the CCSD(T) level.

\color{rev0}
Despite these advances, most current CCSD(T)-level MLIPs still lack application to systems featuring extended covalent networks, such as polymers, metal–organic frameworks (MOFs), and covalent–organic frameworks (COFs).
The primary challenge is the limited availability of CCSD(T) under periodic boundary conditions.
Although explicit periodic implementations have been reported in recent years~\cite{
Booth_N_493_365_2012,
McClain_JCTC_13_1209_2017,
Gruber_PRX_8_021043_2018,
Ye_Misc_2023_Ab,
Ye_FD_254_628_2024,
Ye_JCTC_20_8948_2024,
Herzog_nCM_10_68_2024},
they are not in broad routine use.
To our best knowledge, so far only Herzog~\textit{et al.}~\cite{Herzog_nCM_10_68_2024} employed their own periodic CCSD(T) implementation to train an MLIP for a zeolite.
For systems like molecular crystals and liquid water, fragmentation strategies~\cite{
Hermann_PRL_101_183005_2008,  Yang_S_345_640_2014,  ONeill____2025}  can circumvent explicit periodic CCSD(T) calculations.
However, these approaches are not straightforwardly applicable to genuinely periodic systems with extended covalent networks;
a naive cutting of such systems introduces unpaired valence electrons, and thereby qualitatively different electronic structures.
\color{black}

\color{rev0}
In the present study, we develop a methodology to train MLIPs with CCSD(T) accuracy for systems with extended covalent networks and vdW interactions.
Our methodology is based on the $\Delta$-learning method with a dispersion‑corrected tight‑binding baseline
and an MLIP trained on the differences of the target CCSD(T) energies from the baseline.
This strategy enables the MLIP to be trained solely on molecular systems while maintaining transferability to bulk materials.
\color{black}
We validate the potential by comparing electronic total atomization energies (eTAEs), bond lengths, vibrational frequencies, and intermolecular interaction energies with high‑level reference data, and then apply it to a prototypical COF to analyze its structure, inter‑layer binding energies, and hydrogen absorption at CCSD(T) fidelity.

\section{Methodology}
\label{sec:methods}

\subsection{Quantum-chemical calculations}
\label{sec:methods:molpro}

Quantum-chemical calculations for the training dataset of the MLIP were performed with the MOLPRO~2024.1 program package~\cite{Werner_WCMS_2_242_2012_MOLPRO,Werner_JCP_2020_Molpro}.
The energies were computed by the CCSD(T) method augmented with F12 explicit correlation~\cite{Tenno_JCP_121_117_2004,Tenno_CPL_398_56_2004,Tew_PCCP_9_1921_2007} and employing the pair natural orbital (PNO) based local approximation, i.e., PNO-LCCSD(T)-F12~\cite{Ma_JCTC_14_198_2018_PNO5,Ma_JCTC_17_902_2021_PNO8}.
The correlation-consistent polarized valence triple-$\zeta$ basis sets of Dunning~\cite{Dunning_JCP_1989_Gaussian} augmented with diffuse functions~\cite{Kendall_JCP_1992_Electron} for non-hydrogen atoms, i.e., heavy-aug-cc-pVTZ~\cite{ElSohly_IJQC_2008_Comparison,Papajak_JCTC_2011_Perspectives}, were considered in the present study.
Further details and the rationale behind the applied approaches are described in the following.

The computational cost of the CCSD(T) method in the traditional canonical form scales as $\mathcal{O}(N^7)$, where $N$ is the number of orbitals~\cite{Raghavachari_CPL_157_479_1989}.
This steep scaling of the computational cost limits the application of CCSD(T) to only small molecules, particularly when a large basis set is used.
Local approximations, particularly those based on PNOs~\cite{
Neese_JCP_130_114108_2009,  Neese_JCP_131_064103_2009,  Riplinger_JCP_138_034106_2013,
Riplinger_JCP_139_134101_2013,
Pinski_JCP_143_34108_2015_SparseMaps1,
Riplinger_JCP_144_24109_2016_SparseMaps2,
Guo_JCP_144_94111_2016_SparseMaps3,
Pavosevic_JCP_144_144109_2016_SparseMaps4,
Pavosevic_JCP_146_174108_2017_SparseMaps5,
Guo_JCP_148_011101_2018,
Guo_JCP_152_024116_2020,
Guo_JCP_158_124120_2023_SparseMaps6,
Werner_JCTC_11_484_2015_PNO1,
Ma_JCTC_11_5291_2015_PNO2,
Schwilk_JCTC_13_3650_2017_PNO3,
Ma_JCTC_13_4871_2017_PNO4,*Ma_JCTC_14_6750_2018_PNO4C,
Ma_JCTC_14_198_2018_PNO5,
Krause_JCTC_15_987_2018_PNO6,
Ma_JCTC_16_3135_2020_PNO7,
Ma_JCTC_17_902_2021_PNO8},
dramatically reduce the computational cost and can achieve near-linear scaling.
This enables the computation for systems with a few hundreds of atoms at the CCSD(T) level.

For the F12 correction, which dramatically reduces the basis-set-incompleteness error of the correlation energy~\cite{Tew_PCCP_9_1921_2007}, the F12b approximation~\cite{Adler_JCP_127_221106_2007} with the 3*A ansatz~\cite{Klopper_JCP_116_6397_2002,Werner_JCP_126_164102_2007} and the diagonal fixed amplitude ansatz~\cite{Tenno_JCP_121_117_2004,Tenno_CPL_398_56_2004}
were employed.

The many-electron integrals in the F12 method were approximated by introducing the resolution of the identity (RI) with utilizing the complementary auxiliary basis set (CABS)~\cite{Valeev_CPL_2004_Improving}.
The CABS singles correction~\cite{Adler_JCP_127_221106_2007,Knizia_JCP_128_154103_2008} was considered to reduce the basis-set-incompleteness error of the Hartree--Fock energy.
All electrons including those in the core shells were considered in the correlation methods and the CABS singles correction unless otherwise noted.
The all-electron treatment substantially affects the eTAEs and the vibrational frequencies (\cref{sec:results}), while the impact on dispersion interaction can be marginal (Sec.~\textcolor{violet}{S1} in the Supplemental Materials (SM)).

The basis-set-incompleteness error is typically mitigated by estimating the complete-basis-set (CBS) limit through an extrapolation combining two or more consecutive basis sets from a correlation-consistent family~\cite{Helgaker_JCP_106_9639_1997,Halkier_CPL_286_243_1998}.
However, for local-correlation methods, Werner and Hansen~\cite{Werner_JCTC_19_7007_2023} demonstrated that the basis-set convergence can become non-monotonic because the local approximations introduce basis-set-dependent errors, and they instead recommended using the F12 correction.
Guided by this finding, in the present study, we eschewed the basis-set extrapolation and employed the F12-corrected values obtained with the heavy-aug-cc-pVTZ basis set.

The basis-set-superposition error (BSSE), i.e., the error due to the basis-set-size difference among compared systems, is a particular type of the basis-set-incompleteness error.
The BSSE can be particularly substantial for intermolecular interaction energies.
Specifically, the total energy of a dimer system is not only modified by the interaction of the fragments but also by an improved description of the wavefunction of each fragment due to additional basis functions stemming from the other fragment.
In general, the BSSE may be addressed using, e.g., the counterpoise (CP) correction~\cite{Boys_MP_1970_calculation}.
However, as written above, the F12 method dramatically reduces the basis-set-incompleteness error, which consequently also reduces the BSSE strongly~\cite{McMahon_JCP_135_154309_2011}.
Combined with local-orbital methods, the BSSE is further suppressed~\cite{Saeboe_JCP_98_2170_1993,Schuetz_JPCA_102_5997_1998,GrantHill_PCCP_8_4072_2006,Ma_WCMS_2018_Explicitly}.
These BSSE reductions due to the F12 and the local-orbital approaches are also confirmed in the present study for a benzene--benzene dimer with $\pi$--$\pi$ stacking as detailed in Sec.~\textcolor{violet}{S1} in the SM.
Shortly, in the PNO-LCCSD(T)-F12 method without the CP correction, the errors from the reference data~\cite{Brauer_PCCP_2016_S66x8} are less than \qty{0.6}{kcal/mol}, which is smaller than the chemical-accuracy criterion of \qty{1}{kcal/mol}.
Therefore, in the present study, we did not apply the CP correction and used the raw PNO-LCCSD(T)-F12 values for the sake of simplicity and reduction of computational cost.

The density-fitting approximation was applied to all the methods using the appropriate auxiliary basis sets for approximating Coulomb and exchange contributions to Fock matrix elements~\cite{Weigend_PCCP_4_4285_2002} and two-electron integrals required in the correlation methods~\cite{Weigend_JCP_116_3175_2002}.

For eTAEs, the spin-restricted open-shell variant of PNO-LCCSD(T)-F12, i.e., PNO-RCCSD(T)-F12~\cite{Ma_JCTC_17_902_2021_PNO8} was employed to compute the reference free-atom energies.
The $\rm^2S$  state for H and the $\rm^3P$ state for C were considered.
Note that the H atom is a single-electron system and the Hartree--Fock method already provides the corresponding energy.

For comparison, quantum-chemical calculations were also performed with other methods.
These include the GGA in the Perdew--Burke--Ernzerhof (PBE) form~\cite{Perdew_PRL_77_3865_1996,*Perdew_PRL_78_1396_1997} without and with the D4 vdW correction~\cite{Caldeweyher_JCP_2017_Extension,Caldeweyher_JCP_2019_generally,Caldeweyher_PCCP_2020_Extension}, the second-order Møller--Plesset perturbation theory (MP2), CCSD, and CCSD(T) in the non-PNO form without and with the F12 correction, and the random-phase approximation (RPA) based on PBE Kohn--Sham orbitals.
The heavy-aug-cc-pVTZ basis set was used for these quantum-chemical calculations.

\subsection{MLIP}
\label{sec:methods:MLIP}

\subsubsection{Formalism}

For the MLIP, we utilized the formalism of the moment-tensor potential (MTP)~\cite{Shapeev_MMS_2016_Moment}.
In the MTP formalism, the energy $E^\mathrm{MTP}$ is the sum of contributions $V^\mathrm{MTP}(\mathfrak{n}_i)$ of atomic neighborhoods $\mathfrak{n}_i$ for $N$ atoms;
\begin{align}
E^\mathrm{MTP} = \sum_{i=1}^{N} V^\mathrm{MTP}(\mathfrak{n}_i).
\end{align}
The neighborhood $\mathfrak{n}_i$ is represented as a tuple;
\begin{align}
\mathfrak{n}_i
= (\{r_{i1}, z_{i}, z_{1}\}, \dots, \{r_{ij}, z_{i}, z_{j}\}, \dots, \{r_{iN_\mathrm{nbh}}, z_{i}, z_{N_\mathrm{nbh}}\}),
\end{align}
where $r_{ij}$ are relative atomic positions, $z_i$ and $z_j$ are the types of central and neighboring atoms, and $N_\mathrm{nbh}$ is the number of atoms in neighborhood.
Each contribution $V^\mathrm{MTP}(\mathfrak{n}_i)$ in the potential energy $E^\mathrm{MTP}$ expands through a set of MTP basis functions $B_\alpha$ as
\begin{align}
V^\mathrm{MTP}(\mathfrak{n}_i)
= V_0(z_i) + \sum_{\alpha=1}^{N_\mathrm{lin}}
\xi_\alpha B_\alpha(\mathfrak{n}_i),
\end{align}
where $\xi_\alpha$ are the linear parameters to be optimized, and $N_\mathrm{lin}$ is the number of these parameters.
$V_0$ is the reference free-atom energy.
To define the functional form of the MTP basis functions and the number $N_\mathrm{lin}$, the so-called moment-tensor descriptors are introduced;
\begin{align}
M_{\mu,\nu}(\mathfrak{n}_i)
= \sum_{j=1}^{N_\mathrm{nbh}}
f_\mu(|r_{ij}|, z_i, z_j) \, r_{ij}^{\otimes\mu}.
\end{align}
The angular part is given as the outer product of $r_{ij}$ as
\begin{align}
r_{ij}^{\otimes\nu} \equiv \underbrace{r_{ij} \otimes \dots \otimes r_{ij}}_\textrm{{$\nu$ times}},
\end{align}
and thus a $\nu$-th-order tensor.
The radial part $f_\mu(|r_{ij}|, z_i, z_j)$ is a polynomial function smoothly truncated at the cutoff radius $R_\mathrm{cut}$.
The MTP basis function $B_\alpha$ is a scalar value obtained by tensor contraction of one or more moment-tensor descriptors $M_{\mu,\nu}$.
The number of MTP basis functions is limited by the so-called MTP level, $\mathrm{lev}_\mathrm{max}$.
Specifically, the level of an MTP basis function $B_\alpha$ is defined as
\begin{align}
\mathrm{lev} B_\alpha \equiv \sum_{p=1}^{P} \mathrm{lev} M_{\mu_p, \nu_p},
\end{align}
where $M_{\mu_p,\nu_p}$ are the moment-tensor descriptors involved in $B_\alpha$, and
\begin{align}
\mathrm{lev} M_{\mu,\nu} \equiv 2 + 4\mu + \nu.
\end{align}
For a given $\mathrm{lev}_\mathrm{max}$, only the MTP basis functions with $\mathrm{lev}B_\alpha \le \mathrm{lev}_\mathrm{max}$ are considered.
Further details of the MTP formalism can be found in the literature~\cite{Shapeev_MMS_2016_Moment,Novikov_MLST_2021_MLIP,Podryabinkin_JCP_159_084112_2023}.

The training and evaluation of MTPs were performed using our own implementation~\cite{MOTEP}, which allows fixing the reference free-atom energies $V_0$ to those obtained in the quantum-chemical calculations.
The MTP parameters were optimized in the procedure detailed in Sec.~~\textcolor{violet}{S4} in the SM, where the main optimization method was the Broyden--Fletcher--Goldfarb--Shanno~(BFGS) algorithm as implemented in SciPy~\cite{Virtanen_NM_2020_SciPy} with a maximum number of 1000 iterations.
The loss function optimized was the energy per atom averaged over the configurations in the training dataset.
The cutoff radius $R_\mathrm{cut}$ was set to \qty{7}{\angstrom}.

\subsubsection{\texorpdfstring{$\Delta$}{Δ}-learning}
\label{sec:methods:delta}

\color{rev0}

The major challenge for MLIPs with the CCSD(T)-level accuracy~\cite{
Schran_JCP_148_102310_2017,  Schran_JCTC_14_5068_2018,
Schran_JCTC_16_88_2019,
Topolnicki_JCTC_16_6785_2020,
Schran_JCP_154_51101_2021,
Beckmann_JCTC_18_5492_2022,  Chmiela_NC_9_3887_2018,
Sauceda_JCP_150_114102_2019,
Yu_JPCL_13_5068_2022,  Qu_JCTC_19_3446_2023,  Zhu_JCTC_19_3551_2023,  Daru_PRL_129_226001_2022,
Chen_JCTC_19_4510_2023,
ONeill____2025,
Bogojeski_NC_11_5223_2020,
Herzog_nCM_10_68_2024,
Kim_JACS_147_1042_2024,
Tang_NCS_5_144_2024,*Tang_NCS_5_184_2025,
Smith_NC_2019_Approaching,Smith_SD_2020_ANI}
is the high computational cost of CCSD(T) to prepare sufficiently large training sets.
The $\Delta$-learning method~\cite{Ramakrishnan_JCTC_2015_Big,Meng_TCA_2023_Something} is one of the popular approaches that can circumvent this issue and was employed in several previous studies~\cite{Zhu_MC_9_867_2019,Shuaibi_MLST_2_25007_2020,Wengert_CS_2021_Data,Ruth_JCTC_2022_Machine,Ruth_JCTC_19_4912_2023,Huang_MLST_6_35004_2025,Hashim_Misc_2025_Conformational,ONeill____2025}.
For example, Ruth \textit{et al.}~\cite{Ruth_JCTC_2022_Machine,Ruth_JCTC_19_4912_2023} employed the $\Delta$-learning method to predict the energies of small organic molecular systems in their ground-state structures at the CCSD(T) level with the baseline of lower-accuracy methods including DFT. O’Neill \textit{et al.}~\cite{ONeill____2025} recently applied the $\Delta$-learning method to develop CCSD(T)-accuracy MLIPs for liquid water with the DFT baseline, which are available for molecular dynamics (MD) simulations at finite temperatures.

In the present study, the $\Delta$-learning method is applied
with the baseline of the tight-binding method, specifically the GFN2-xTB method~\cite{Bannwarth_JCTC_2019_GFN2}.
The energies of the molecular configurations in the training dataset are thus decomposed as
\color{black}
\begin{align}
E(\textrm{PNO-CCSD(T)-F12}) = E(\textrm{GFN2-xTB}) + \Delta E,
\label{eq:delta}
\end{align}
where $E(\textrm{PNO-CCSD(T)-F12})$ and $E(\textrm{GFN2-xTB})$ are the energies with the PNO-CCSD(T)-F12 and the GFN2-xTB methods, respectively, and $\Delta E$ is the energy difference between the two methods.
An MTP is fitted to $\Delta E$ based on the energies of the molecular systems in the training dataset.
By summing up the energies or the forces of GFN2-xTB and the trained MTP, we reproduce the potential energy surface (PES) of the PNO-CCSD(T)-F12 method.
We hereafter refer to the MTP trained on the $\Delta E$ as the $\Delta$MTP and the sum of GFN2-xTB and $\Delta$MTP as TB+$\Delta$MTP.

In the above $\Delta$-learning approach, it is supposed that major features of the PES are already well reproduced in GFN2-xTB and that $\Delta E$ provides corrections.
A smaller number of configurations are therefore required to train an MTP for $\Delta E$ than for $E(\textrm{PNO-CCSD(T)-F12})$ directly.
\textcolor{rev0}{It is also assumed that the energetic differences between GFN2-xTB and CCSD(T) are predominantly local.
This allows us to train the $\Delta$MTP that is available even for periodic systems such as COFs, where CCSD(T) is not yet routinely available.
In other words, the training dataset for $\Delta E$ can still be constructed from the molecular systems that represent local regions of the target periodic systems that have extended covalent networks.}
Also, since GFN2-xTB is several orders faster than DFT, and an MLIP is even faster than GFN2-xTB, the $\Delta$-learning approach makes it possible to evaluate the energies of the target system essentially only with the tight-binding cost.
Another advantage of taking GFN2-xTB as the reference is that
London dispersion interactions are already approximated with good accuracy by the D4 formalism~\cite{Caldeweyher_JCP_2019_generally}.
As demonstrated below for a benzene--benzene dimer with $\pi$--$\pi$ stacking (\cref{sec:results:dimer}), the London dispersion interaction \textcolor{rev0}{at large distances} is well reproduced already in GFN2-xTB, and therefore only the region around the equilibrium distance needs to be corrected.
Therefore, the \textcolor{rev0}{$\Delta$}MTP needs to be trained only with a moderate cutoff radius, which can also reduce the sizes of the molecular systems in the training dataset.

The GFN2-xTB calculations were performed with the TBLITE code~\footnote{\url{https://tblite.readthedocs.io}} in combination with the interface for the ASE code~\cite{Larsen_JPCM_2017_atomic}.
For the calculations of free atoms, an extension for open-shell high-spin states, i.e., spGFN2-xTB~\cite{Neugebauer_JCC_44_2120_2023} was employed.

\subsubsection{Training, validation, and test datasets}

\begin{figure*}[!tb]
\centering
\includegraphics{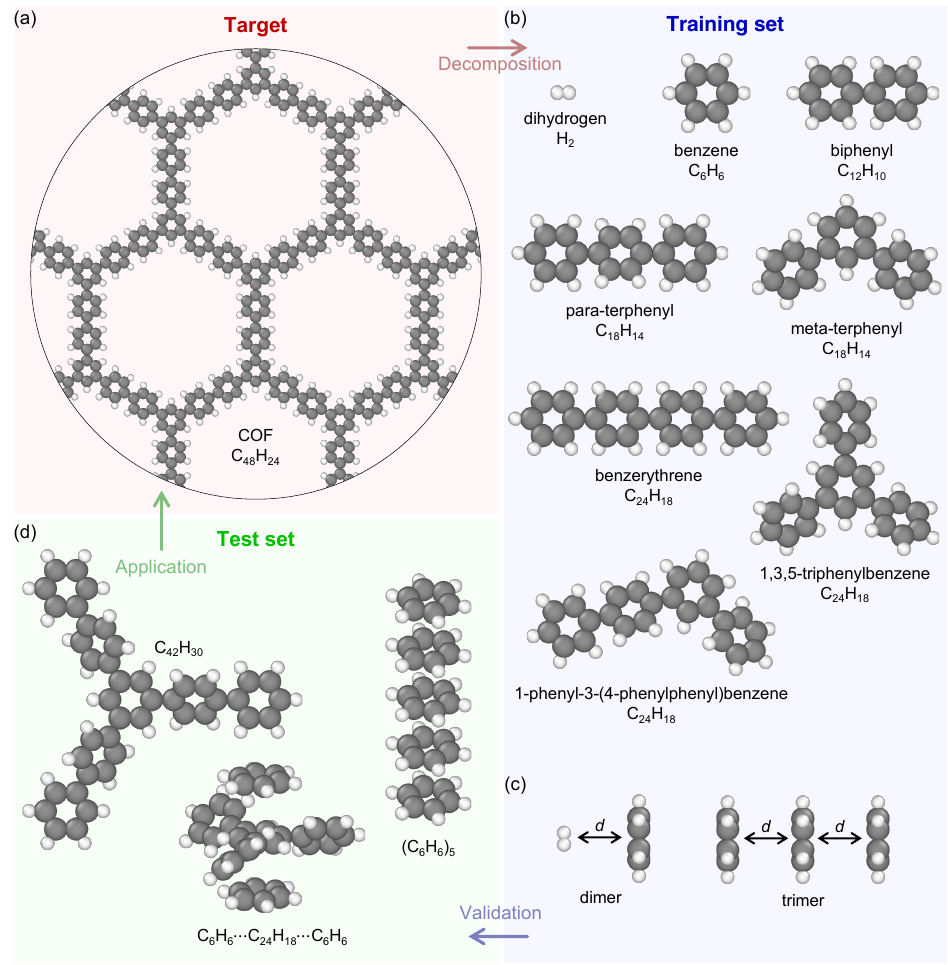}
\caption{(a) Single layer of the periodic \ce{C48H30} COF.
(b) Monomer molecules considered for the training of MTPs.
(c) Dihydrogen--benzene dimer and benzene trimer with the center-of-mass distances~$d$ as examples of multimers considered also for the training.
(d) Molecular systems used for testing the trained MTPs, thus not included in the training datasets.
Visualization was performed using OVITO~\cite{Stukowski_MSMSE_2010_Visualization}.}
\label{fig:molecules}
\end{figure*}

COFs~\cite{Cote_S_2005_Porous} are nanoporous crystalline materials formed by covalent bonds of organic secondary building units composed mainly of light elements like C, N, O, as well as H.
The covalent-bond networks are extended either in a two-dimensional (2D) or a three-dimensional (3D) way.
The majority of experimentally synthesized COFs are quasi-2D layered materials~\cite{Tong_CES_2017_Exploring,Tong_JPCC_2018_Computation,Ongari_ACS_2019_Building} held together by vdW interactions.
Due to their light weight, high porosity, and chemical diversity,
COFs have attracted great interest for various applications, such as the storage and separation of gases like hydrogen (H\textsubscript{2}), methane (CH\textsubscript{4}), carbon dioxide (CO\textsubscript{2}), and ammonia (NH\textsubscript{3})~\cite{
Han_JACS_2008_Covalent,  Furukawa_JACS_2009_Storage,  Doonan_NC_2010_Exceptional},  water harvesting~\cite{Stegbauer_CM_2015_Tunable,Nguyen_JACS_2020_Porous,Grunenberg_JACS_2023_Postsynthetic},
and photocatalysis for hydrogen evolution from water~\cite{Vyas_NC_2015_tunable,Wang_NC_2018_Sulfone,Biswal_JotACS_2019_Sustained}.

To demonstrate the applicability of the developed $\Delta$-learning approach, we considered a prototypical COF consisting of carbon and hydrogen synthesized in experiments on the Au(111) surface~\cite{Blunt_CC_46_7157_2010,Shi_CC_52_8726_2016}.
This COF has a quasi-2D structure with each layer being a network of benzene rings.
The 2D unit cell of a single layer is hexagonal, and the composition of each layer in the unit cell is \ce{C48H30}.
We therefore hereafter refer to this COF as the \ce{C48H30} COF.
This COF is labeled 10020N2 or 16470N2 (both having the topologically identical structure) in the CURATED-COFs database~\cite{Ongari_ACS_2019_Building}.
\cref{fig:molecules}(a) presents a single layer of the \ce{C48H30} COF.

\begin{table*}[tb]
\centering
\caption{Configurations in the training, validation, and test datasets.\footnote{$N_\mathrm{ring}$: Number of benzene rings per isomer.
$N_\mathrm{isomer}$: Number of isomers considered.
$N_\mathrm{dist}$: Number of considered intermolecular distances.
$N_\mathrm{conf}^\mathrm{train}$: Number of configurations in the training dataset.
$N_\mathrm{conf}^\mathrm{valid}$: Number of configurations in the validation set.
$N_\mathrm{conf}^\mathrm{test}$: Number of configuration in the test dataset.}}
\label{tab:multimers}
\begin{tabular}{l*1{l}*{2}{c}*{4}{r}}
\toprule
&
\multicolumn{1}{c}{System} &
\multicolumn{1}{c}{$N_\mathrm{ring}$} &
\multicolumn{1}{c}{$N_\mathrm{isomer}$} &
\multicolumn{1}{c}{$N_\mathrm{dist}$} &
\multicolumn{1}{c}{$N_\mathrm{conf}^\mathrm{train}$} &
\multicolumn{1}{c}{$N_\mathrm{conf}^\mathrm{valid}$} &
\multicolumn{1}{c}{$N_\mathrm{conf}^\mathrm{test }$} \\
\midrule
Monomer
& \ce{H2}     & 0 & 1 & n/a &   9 &   1 &   0 \\
& \ce{C6H6}   & 1 & 1 & n/a &   9 &   1 &   0 \\
& \ce{C12H10} & 2 & 1 & n/a &   9 &   1 &   0 \\
& \ce{C18H14} & 3 & 2 & n/a &  18 &   2 &   0 \\
& \ce{C24H18} & 4 & 3 & n/a &  27 &   3 &   0 \\
& \ce{C42H30} & 7 & 1 & n/a &   0 &   0 &  11 \\
\midrule
Dimer
& \ce{H2\bond{...}H2}         & 0 & 1 & 10 &  90 &  10 & 0 \\
& \ce{H2\bond{...}C6H6}       & 1 & 1 & 10 &  90 &  10 & 0 \\
& \ce{H2\bond{...}C12H10}     & 2 & 1 & 10 &  90 &  10 & 0 \\
& \ce{H2\bond{...}C18H14}     & 3 & 2 & 10 & 180 &  20 & 0 \\
& \ce{C6H6\bond{...}C6H6}     & 2 & 1 & 10 &  90 &  10 & 0 \\
& \ce{C6H6\bond{...}C12H10}   & 3 & 1 & 10 &  90 &  10 & 0 \\
& \ce{C6H6\bond{...}C18H14}   & 4 & 2 & 10 & 180 &  20 & 0 \\
& \ce{C6H6\bond{...}C24H18}   & 5 & 3 & 10 & 270 &  30 & 0 \\
& \ce{C12H10\bond{...}C12H10} & 4 & 1 & 10 &  90 &  10 & 0 \\
\midrule
Trimer
& \ce{H2\bond{...}H2\bond{...}H2}         & 0 & 1 & 10 &  90 & 10 &  0 \\
& \ce{H2\bond{...}C6H6\bond{...}H2}       & 1 & 1 & 10 &  90 & 10 &  0 \\
& \ce{H2\bond{...}C12H10\bond{...}H2}     & 2 & 1 & 10 &  90 & 10 &  0 \\
& \ce{C6H6\bond{...}H2\bond{...}C6H6}     & 2 & 1 & 10 &  90 & 10 &  0 \\
& \ce{C6H6\bond{...}C6H6\bond{...}C6H6}   & 3 & 1 & 10 &  90 & 10 &  0 \\
& \ce{C6H6\bond{...}C12H10\bond{...}C6H6} & 4 & 1 & 10 &  90 & 10 &  0 \\
& \ce{C6H6\bond{...}C24H18\bond{...}C6H6} & 6 & 1 &  8 &   0 &  0 & 88 \\
\midrule
Tetramer
& \ce{C6H6\bond{...}C6H6\bond{...}C6H6\bond{...}C6H6} & 4 & 1 & 10 & 90 & 10 & 0 \\
\midrule
Pentamer
& \ce{C6H6\bond{...}C6H6\bond{...}C6H6\bond{...}C6H6\bond{...}C6H6} & 5 & 1 & 8 & 0 & 0 & 88 \\
\bottomrule
\end{tabular} \end{table*}

The molecular systems for the training of the MTPs were selected from ``components'' of the target periodic COF, i.e., those that can be made by decomposing \textcolor{rev0}{the COF into small fragments, with hydrogen termination to avoid unpaired valence electrons}.
\cref{fig:molecules}(b) shows the thus considered monomer molecules.
In order to study also hydrogen absorption, dihydrogen (\ce{H2}) was also included.
To capture the inter-layer vdW interaction, several dimers, trimers, and tetramers composed of these monomers were also included in the training set, as exemplified in \cref{fig:molecules}(c).
In order to cover a wide range of inter-molecular distance on an equal footing, ten center-of-mass distances between each monomer were considered.
For each molecular system, MD simulations (detailed below) were conducted, and ten snapshots in the MD trajectory were selected.
Nine among them were cumulated into the master training dataset, while the last one was reserved for the validation of the MTPs after the training, which we hereafter refer to as the validation dataset.
\cref{tab:multimers} summarizes the considered molecular systems together with the numbers of configurations in both the datasets.

To assess the effect of the system size in the training dataset on MTP quality, four training sets were constructed from the master training dataset above based on the maximum number of benzene rings permitted in the systems. The first set included only molecules with up to two benzene rings, and the resulting $\Delta$MTP is labeled~\#2. Analogously, three more training sets allowed up to three, four, and five benzene rings, yielding models \#3, \#4, and \#5, respectively.
Specifically, $\Delta$MTP\#5 included 8 monomers, 12 dimers, 6 trimers, and 1 tetramer, containing at maximum 30 carbon atoms.

In addition to the molecular systems in the training datasets, we also evaluated larger molecular systems as test datasets to examine whether the trained MTPs retain predictive accuracy for systems beyond those used in training, thereby indirectly supporting their transferability to the periodic \ce{C48H24} COF.
The systems for the test dataset consist of a 1,3,5-tri(4-biphenyl)benzene monomer (\ce{C42H30}), a benzene--1,3,5-triphenylbenzene--benzene trimer (\ce{C6H6\bond{...}C24H18\bond{...}C6H6}), and a benzene pentamer (\ce{(C6H6)5}), as presented in \cref{fig:molecules}(d).

The datasets of MLIPs for finite-temperature simulations are often made by MD simulations~\cite{
Forslund_PRB_2021_Ab,
Grabowski_nCM_2019_Ab,
Gubaev_PRM_2021_Finite,
Zhu_CMS_2021_fully,  Novikov_nCM_2022_Magnetic,
Zhou_PRB_2022_Thermodynamics,
Forslund_PRB_2023_Thermodynamic,
Gubaev_nCM_9_129_2023,
Jung_nCM_2023_High,
Jung_PRB_2023_Dynamically,
Srinivasan_PRB_2023_Anharmonicity,
Xu_AM_2023_Strong,
Ou_PRM_8_115407_2024,
Srinivasan_nCM_2024_Electronic,
Xu_AM_2024_Origin,
Xu_SM_2024_Accurate,
Zhu_PRB_2024_Melting,
Zhu_nCM_2024_Accelerating,
Zotov_MSMS_2024_Moment,
Zhang_NC_16_394_2025},
the strategy adopted also in the present study.
As MD simulations are prohibitively expensive at the coupled-cluster level, we instead used GFN2-xTB for MD simulations.
For selected snapshots from these trajectories, PNO-LCCSD(T)-F12 single-point energies were computed.

Further details of the geometry generation are as follows.
The geometries of the monomers were first obtained from the ChemSpider website~\cite{Pence_JCE_87_1123_2010}.
They were then relaxed with the GFN2-xTB method using the BFGS algorithm implemented in ASE~\cite{Larsen_JPCM_2017_atomic} until all the forces on atoms became less than \qty{0.001}{eV/\angstrom}.
The MD simulations were then performed starting from the relaxed geometries in the Langevin thermostat implemented in ASE at \qty{300}{K} with a friction parameter of \qty{0.01}{fs\textsuperscript{$-$1}} for a simulation time of \qty{1}{ps} with a time step of \qty{1}{fs}, thus 1001~steps including the zeroth step.
For multimers, their constituent monomers were initially aligned along the axis corresponding to the largest principal moment of inertia, and the center-of-mass distances of the neighboring monomers were fixed to a given value during the MD simulations.
Ten values in \qtyrange{3.0}{7.5}{\angstrom} with a step of \qty{0.5}{\angstrom} were considered as the center-of-mass distances for the training and the validation datasets, while eight values in \qtyrange{4.0}{7.5}{\angstrom} were considered for the test dataset.
(Shorter distances were excluded because quantum-chemical calculations often raised errors due to basis-set linear dependence.)
Thus, specifically for $\Delta$MTP\#5, 1872 and 208, and 187 configurations were in the training, validation, and test test datasets.

\subsubsection{Local extrapolation grade}

The trained $\Delta$MTPs were also evaluated based on their uncertainties for given configurations.
Specifically, we employed the extrapolation grade~\cite{Podryabinkin_CMS_2017_Active,Gubaev_CMS_2019_Accelerating} defined based on D-optimality.
The extrapolation grade evaluates the similarity of the given configurations or the local atomic environments to those in the training dataset.
A value less than~1 indicates that the considered configuration of the local environment is interpolative and that the prediction should be reliable, while a higher extrapolation grade implies a larger uncertainty for properties predicted by the potential.
The extrapolation grade is advantageous because this can be evaluated only based on geometry and thus for periodic systems.
The extrapolation grade has often been employed as the uncertainty measure during active learning~\cite{Gubaev_PRM_2021_Finite,
Lysogorskiy_PRM_2023_Active,
Ou_PRM_8_115407_2024,
Kumar_AM_297_121319_2025}.
As applied for high-entropy alloys in our previous study~\cite{Gubaev_PRM_2021_Finite}, an extrapolation grade less than~2 should be safe for application.

The extrapolation grade can be defined both for a whole configuration and for each atom~\cite{Podryabinkin_CMS_2017_Active,Gubaev_CMS_2019_Accelerating}, and in the present study, we considered the latter one, which we hereafter refer to as the local extrapolation grade.
The MLIP3 package~\cite{Podryabinkin_JCP_159_084112_2023} was employed for the calculation of the local extrapolation grade.
The local extrapolation grades were evaluated for the molecular systems in the test dataset and for the periodic \ce{C48H30} COF.

\subsubsection{Comparison with ANI-1ccx}

The performance of TB+$\Delta$MTP was compared with another MLIP, ANI-1ccx~\cite{Smith_NC_2019_Approaching,Smith_SD_2020_ANI}.
The ANI-1ccx potential is an ANI-style neural network potential~\cite{Smith_CS_2017_ANI} trained on the CCSD(T)-level training dataset aiming for broad organic chemistry applications.
As many as \num{500000} configurations of molecules consisting of the elements in \{H, C, N, O\} were employed for the training, while \ce{H2} is absent in the dataset.
The CCSD(T)-level calculations were performed with the cc-pV\textit{X}Z basis sets up to with \textit{X} in \{D, T\}, without augmenting diffuse functions.
The cutoff radius was set to~\qty{5.2}{\angstrom}.
Free-atom energies were obtained by linear fitting to the energy per atomic species in the training dataset.
The ANI-1ccx calculations were performed with the TorchANI code~\cite{Gao_JCIM_2020_TorchANI} with the interface for the ASE code~\cite{Larsen_JPCM_2017_atomic}.

\section{Results and Discussion}
\label{sec:results}

\subsection{RMSEs of MTPs}
\label{sec:results:MTPs}

\begin{figure*}[tbp]
\centering
\includegraphics{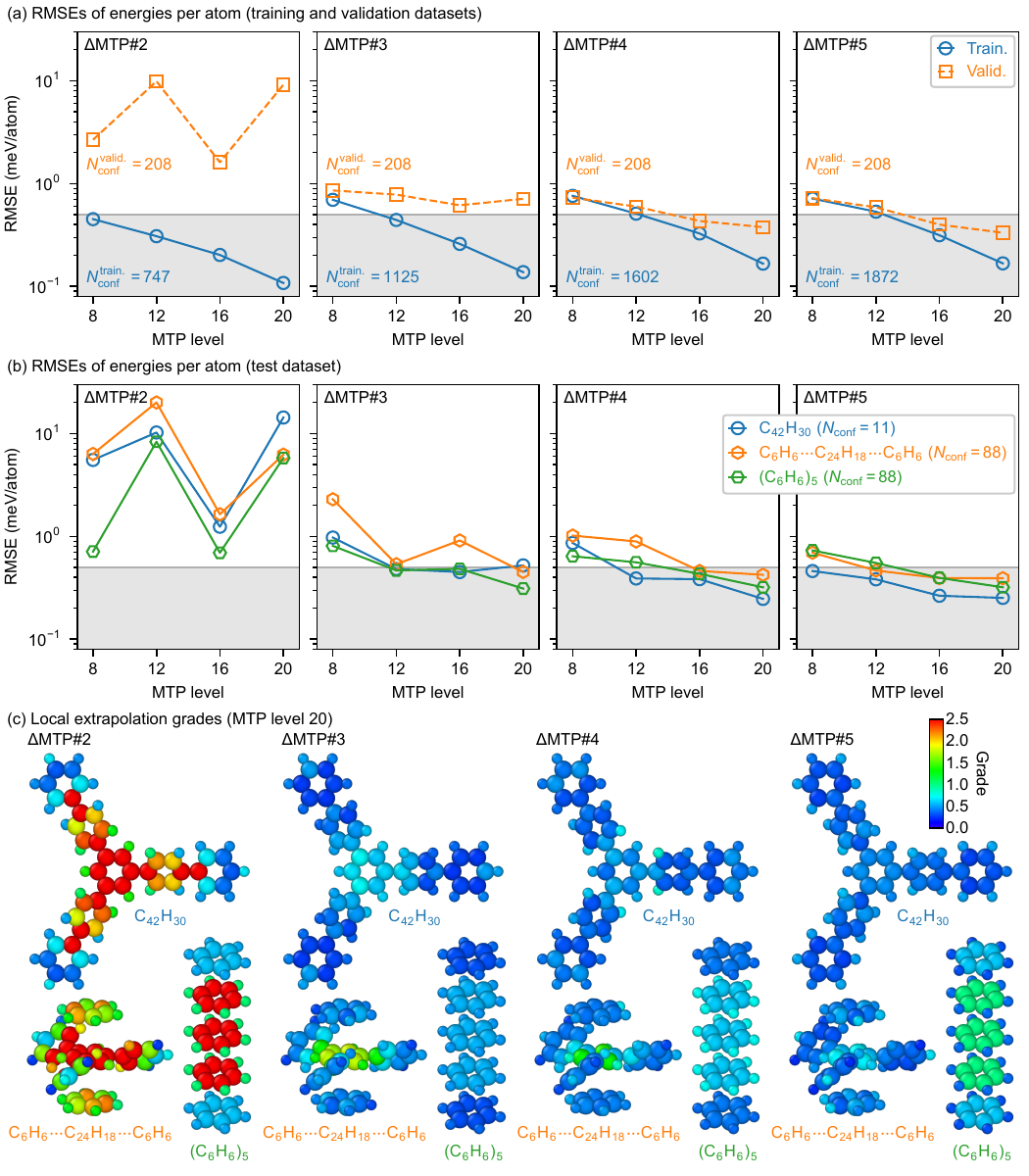}
\caption{(a) RMSEs of the energies per atom for the molecular systems in the training and the validation datasets predicted by $\Delta$MTPs.
(b) RMSEs of the energies per atom for the molecular systems in the test dataset.
(c) Local extrapolation grades of a selected snapshot from each molecular system in the test dataset obtained with $\Delta$MTPs at level 20.}
\label{fig:validation}
\end{figure*}

\begin{figure}[!tb]
    \centering
    \includegraphics{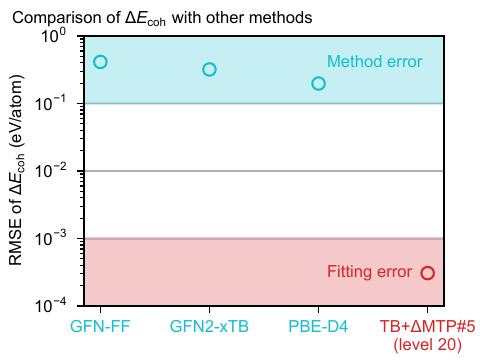}
    \caption{RMSEs of cohesive energies per atom for the training datasets in GFN-FF~\cite{Spicher_ACIE_59_15665_2020}, GFN2-xTB~\cite{Bannwarth_JCTC_2019_GFN2}, PBE-D4~\cite{Perdew_PRL_77_3865_1996,*Perdew_PRL_78_1396_1997,Caldeweyher_JCP_2019_generally}, as well as TB+$\Delta$MTP\#5 at the MTP level 20 with respect to the reference PNO-LCCSD(T)-F12 values.}
    \label{fig:cohesive_energies_vs_methods}
\end{figure}

\begin{table}[!tb]
\centering
\caption{RMSEs (in meV/atom) for the investigated datasets obtained with $\Delta$MTP\#\textit{N}.}
\label{tab:RMSEs}
\begin{tabular}{rr*{3}{S[table-format=4.3]}}
\toprule
\multicolumn{1}{c}{\textit{N}} &
\multicolumn{1}{c}{Level} &
\multicolumn{1}{c}{Training\#\textit{N}} &
\multicolumn{1}{c}{Validation} &
\multicolumn{1}{c}{Test} \\
\midrule
2
 &  8 & 0.451 & 2.671 &  4.577 \\
 & 12 & 0.308 & 9.888 & 15.119 \\
 & 16 & 0.202 & 1.613 &  1.259 \\
 & 20 & 0.108 & 9.147 &  6.797 \\
\midrule
3
 &  8 & 0.695 & 0.857 &  1.687 \\
 & 12 & 0.443 & 0.781 &  0.501 \\
 & 16 & 0.259 & 0.615 &  0.717 \\
 & 20 & 0.138 & 0.712 &  0.397 \\
\midrule
4
 &  8 & 0.760 & 0.730 &  0.852 \\
 & 12 & 0.510 & 0.598 &  0.729 \\
 & 16 & 0.327 & 0.433 &  0.444 \\
 & 20 & 0.166 & 0.377 &  0.368 \\
\midrule
5
 &  8 & 0.715 & 0.721 &  0.700 \\
 & 12 & 0.533 & 0.587 &  0.504 \\
 & 16 & 0.315 & 0.399 &  0.387 \\
 & 20 & 0.168 & 0.333 &  0.352 \\
\bottomrule
\end{tabular}
\end{table}

\cref{fig:validation}(a) shows the root-mean-square errors (RMSEs) of the energies per atom for the molecular systems in the training and the validation datasets predicted by $\Delta$MTPs for several MTP levels.
\cref{tab:RMSEs} summarizes the RMSEs for the datasets obtained with the $\Delta$MTPs.
For the training datasets, all the $\Delta$MTPs show monotonous decreases of RMSEs with increasing MTP level, i.e., number of parameters.
For the validation dataset, in contrast, $\Delta$MTP\#2 shows about one order larger RMSEs than those for the training dataset.
This implies overfitting due to less configurations in the corresponding training dataset with respect to the number of MTP parameters as well as due to the incapability to predict larger molecules in the validation dataset not included in the training dataset.
With increasing the sizes of molecular systems in the training dataset, the difference between the RMSEs for the training and the validation datasets becomes smaller, demonstrating the suppression of the overfitting above.
Particularly for $\Delta$MTP\#4 and $\Delta$MTP\#5 at MTP levels of 16 and 20, the RMSEs for the validation dataset are about \qty{0.4}{meV/atom} or lower.
For these $\Delta$MTPs, however, while the RMSEs for the validation dataset decrease monotonously until an MTP level of 16, the RMSEs of the MTP levels of 16 and 20 are almost the same.
This implies that a further higher MTP level would not further improve the transferability of the MTP due to overfitting.
The test dataset also shows similar RMSEs as the validation datasets, as found in \cref{fig:validation}(b), indicating that the predictive power is preserved even for the extended molecular systems not included in the training datasets.

\cref{fig:validation}(c) visualizes the local extrapolation grades of selected configurations in the test dataset obtained with the $\Delta$MTPs.
The extrapolation grades tend to be lower for the atoms in the exterior region of each system, and those for the atoms in the interior region tend to be higher.
Particularly for $\Delta$MTP\#2, the extrapolation grades in the interior regions are higher than 2.5 and thus indicate large uncertainty for this $\Delta$MTP.
In contrast, for $\Delta$MTPs with larger training datasets, the extrapolation grades are much lower, supporting the transferability of these MTPs even for the extended systems not included in the training datasets.
The lowering of the extrapolation grades are found not only for the monomer molecule (\ce{C42H30}) but also for the benzene pentamer (\ce{(C6H6)5}), meaning that the uncertainty decreases not only for covalent interactions but also for vdW interactions.
This improvement demonstrates that the strategy to apply local MLIPs to extended systems should work in combination with the $\Delta$-learning approach with the GFN2-xTB baseline.

\cref{fig:cohesive_energies_vs_methods} shows the differences of the cohesive energies per atom for the \#5 training dataset computed with various methods with respect to the reference PNO-LCCSD(T)-F12 values.
The errors due to the MTP fitting are three orders smaller than the errors due to different methods, e.g., PBE-D4, indicating that TB+$\Delta$MTP\#5 predicts the CCSD(T)-level energies much more accurately than the other methods.

Unless otherwise stated, we hereafter focus on the results with TB+$\Delta$MTP\#5 at level 20.

\subsection{Electronic total atomization energies of \texorpdfstring{H\textsubscript{2}}{H2} and \texorpdfstring{C\textsubscript{6}H\textsubscript{6}}{C6H6}}
\label{sec:results:atomization_energies}

\begin{table}[!tbp]
\small
\centering
\caption{Electronic total atomization energy (kcal/mol) of~\ce{H2}.\footnote{The heavy-aug-cc-pVTZ basis set was used for quantum-chemical calculations.}}
\label{tab:atomization_energies_H2}
\begin{tabular}{lS[table-format=5.1,table-auto-round]}
\toprule
\multicolumn{1}{c}{Method} &
\multicolumn{1}{c}{eTAE} \\
\midrule
GFN2-xTB                   &  77.8 \\  \midrule
PBE                        & 104.7 \\
PBE-D4                     & 104.8 \\
\midrule
PNO-LCCSD-F12\footnotemark[2]& 109.18489471351275 \\
\midrule
ANI-1ccx\footnotemark[3]   &   1.0 \\
\bfTBDMTP[5]               & 109.188311780 \\
\midrule
Exp.\footnotemark[4]       & 109.5 \\
\bottomrule
\end{tabular} \footnotetext[2]{\ce{H2} has only two electrons and therefore no triple excitations.}
\footnotetext[3]{Free-atom energies were obtained by linear fitting to the energy per atomic species in the training dataset.}
\footnotetext[4]{Huber and Herzberg~\cite{Huber_Book_1979_Molecular} with subtracting the vibrational zero-point energy from Irikura~\cite{Irikura_JPCRD_2007_Experimental,*Irikura_JPCRD_2009_Erratum}.}
\end{table}

\begin{table}[!tb]
\small
\centering
\caption{Electronic total atomization energy (kcal/mol) of~\ce{C6H6}.\footnote[1]{The heavy-aug-cc-pVTZ basis set was used for quantum-chemical calculations.}}
\label{tab:atomization_energies_C6H6}
\begin{tabular}{lcS[table-format=4.3]}
\toprule
\multicolumn{1}{c}{Method} &
Correlation &
\multicolumn{1}{c}{eTAE} \\
\midrule
GFN2-xTB                   & n/a    & 1425.6 \\  \midrule
PBE                        & n/a    & 1414.1 \\
PBE-D4                     & n/a    & 1420.8 \\
\midrule
PNO-LCCSD(T)-F12           & frozen-core & 1360.9 \\
                           & all-electron & 1366.5 \\
\midrule
ANI-1ccx\footnotemark[2]   & n/a    &   39.4 \\
\bfTBDMTP[5]               & n/a    & 1366.5 \\
\midrule
Exp.\footnotemark[3]       & n/a    & 1367.8(7) \\
\bottomrule
\end{tabular} \footnotetext[2]{Free-atom energies were obtained by linear fitting to the energy per atomic species in the training dataset.}
\footnotetext[3]{Estimation of Parthiban and Martin, which intrinsically contains scalar relativistic effects (obtained as \qty{-0.99}{kcal/mol}) and spin-orbit coupling (\qty{-0.51}{kcal/mol})~\cite{Parthiban_JCP_2001_Fully}.}
\end{table}

\cref{tab:atomization_energies_H2} shows the eTAEs of dihydrogen (\ce{H2}) obtained by (sp)GFN2-xTB, various quantum-chemical methods, ANI-1ccx, and TB+$\Delta$MTP\#5, as well as the experimental values (where the contributions of zero-point vibrations are subtracted)~\cite{Huber_Book_1979_Molecular,Irikura_JPCRD_2007_Experimental,*Irikura_JPCRD_2009_Erratum}.
The GFN2-xTB method underestimates the TAEs of \ce{H2} by \qty{31.7}{kcal/mol}.
The errors are largely reduced in DFT.
For PNO-LCCSD-F12, the deviation from the experimental reference is only \qty{0.3}{kcal/mol}, well below the chemical-accuracy criterion of \qty{1}{kcal/mol}.
(Note that \ce{H2} is a two-electron system, and therefore there is no triples correction.)
TB+$\Delta$MTP\#5 essentially perfectly reproduces the target PNO-CCSD-F12 value.
In contrast, ANI-1ccx shows a huge underestimation of the TAE, likely because its training dataset does not include \ce{H2}.

\cref{tab:atomization_energies_C6H6} shows the eTAEs of benzene (\ce{C6H6}) computed in various approaches.
The (sp)GFN2-xTB method shows an overestimation by \qty{+57.8}{kcal/mol} from the experimental value~\cite{Parthiban_JCP_2001_Fully}.
The errors are still large in the DFT methods; for example, PBE shows an error of \qty{+47.3}{kcal/mol}, far beyond the chemical accuracy.
The D4 dispersion correction makes the deviation even larger.

The all-electron correlation treatment is found essential for accurate calculations of TAEs for PNO-LCCSD(T)-F12.
In the frozen-core approximation, PNO-LCCSD(T)-F12 shows a derivation of \qty{-6.9}{kcal/mol} from experiments, while the error becomes much smaller in the all-electron treatment, \qty{-1.3}{kcal/mol}, near the chemical accuracy.
A similar impact of all-electron treatment (\qty{7.1}{kcal/mol}) was found also in an accurate coupled-cluster level \textit{ab initio} calculation by Parthiban and Martin~\cite{Parthiban_JCP_2001_Fully}.

TB+$\Delta$MTP\#5 again essentially perfectly reproduces the target value of PNO-LCCSD(T)-F12 and thus yields the same small deviation from experiments.
In contrast, ANI-1ccx shows a large error of \qty{-1328.4}{kcal/mol}.
This is likely because free-atom energies in ANI-1ccx are determined by linear fitting over the entire dataset of molecules rather than by actual quantum-chemical calculations.

\subsection{Bond lengths of \texorpdfstring{H\textsubscript{2}}{H2} and \texorpdfstring{C\textsubscript{6}H\textsubscript{6}}{C6H6}}
\label{sec:results:bond_lengths}

\begin{figure}[!tb]
\centering
\includegraphics{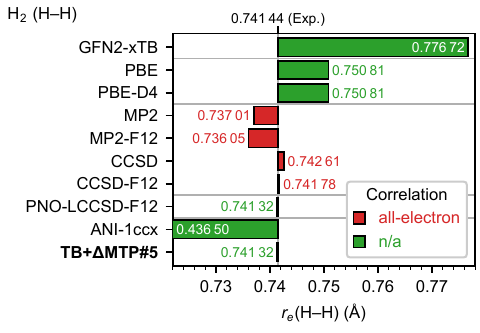}
\label{fig:bond_lengths_H2}
\caption{Equilibrium bond length (\unit{\angstrom}) of dihydrogen (\ce{H2}) with respect to the experimental value in Huber and Herzberg~\cite{Huber_Book_1979_Molecular}.
The heavy-aug-cc-pVTZ basis set was used for quantum-chemical calculations.
Note that \ce{H2} has only two electrons and therefore no triple excitations.
The PNO-LCCSD(T)-F12 value was obtained by fitting the energies for H--H bond lengths on a \qty{0.001}{\angstrom} grid to an univariate second-order polynomial.}
\end{figure}

\cref{fig:bond_lengths_H2} presents the equilibrium bond lengths of dihydrogen (\ce{H2}) obtained using GFN2-xTB, several quantum-chemical methods, ANI-1ccx, and TB+$\Delta$MTP\#5, as well as the value derived from experiments (without contributions of zero-point vibrations)~\cite{Huber_Book_1979_Molecular}.
The GFN2-xTB method leads to an overestimation of about \qty{0.04}{\angstrom} relative to the experimental value.
The DFT methods show substantially smaller deviations by approximately \qty{0.01}{\angstrom}.
Notably, the PBE-D4 vdW functional does not improve the bond length over the PBE result.
The post-HF methods show even smaller deviations from experiments;
particularly, PNO-LCCSD-F12 shows a deviation as small as \qty{0.0001}{\angstrom}, indicating the accuracy of the method.
TB+$\Delta$MTP\#5 reproduces the target value of PNO-LCCSD-F12 almost perfectly, with a deviation of less than \qty{0.00001}{\angstrom}.
ANI-1ccx does not reproduce the H--H bond length even qualitatively, again because the training dataset for this potential does not include \ce{H2}.

\begin{figure}[tb!]
\centering
\includegraphics{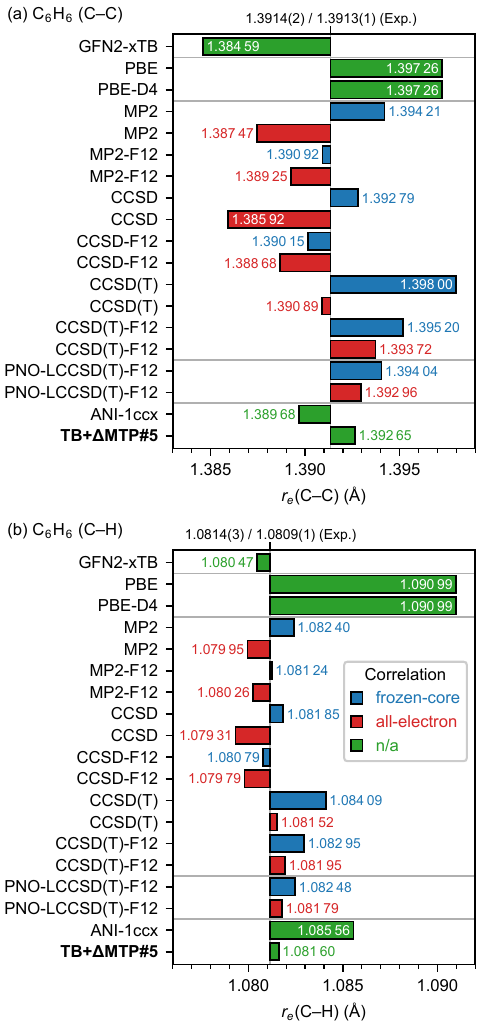}
\label{fig:bond_lengths_C6H6}
\caption{Equilibrium bond lengths (\unit{\angstrom}) of benzene (\ce{C6H6}) with respect to the experimental values of Heo~\textit{et al.}~\cite{Heo_JPCL_2022_Mass} and Esselman~\textit{et al.}~\cite{Esselman_JACS_2023_Precise}.
The heavy-aug-cc-pVTZ basis set was used for quantum-chemical calculations.
The PNO-LCCSD(T)-F12 values were obtained by fitting the energies for C--C and C--H bond lengths on a \qty{0.001}{\angstrom} grid to a bivariate second-order polynomial.}
\end{figure}

\cref{fig:bond_lengths_C6H6} presents the equilibrium bond lengths of benzene (\ce{C6H6}) obtained using GFN2-xTB, several quantum-chemical methods, ANI-1ccx, and TB+$\Delta$MTP\#5, as well as the values derived from experiments (without contributions of zero-point vibrations)~\cite{Heo_JPCL_2022_Mass,Esselman_JACS_2023_Precise}.

The DFT values deviate from experiments by less than \qty{0.006}{\angstrom} for the C--C bonds but as large as \qty{0.01}{\angstrom} for the C--H bonds.
PBE-D4 exhibits deviations virtually identical to those of plain PBE, as with \ce{H2} above.

Regarding post-HF methods, it is worth first discussing the relation between all-electron treatment and F12 explicit inter-electronic correlation.
Using the frozen‑core approximation, MP2, CCSD, and CCSD(T) overestimate C–C bond lengths by about \qty{0.007}{\angstrom} and C–H bond lengths by about \qty{0.002}{\angstrom} relative to all‑electron values.
Adding the F12 explicit‑electron‑correlation method reduces these errors to roughly \qtyrange{0.001}{0.002}{\angstrom}.
Apparent discrepancies between frozen‑core and all‑electron treatments in non‑F12 calculations likely arise from different rates of basis‑set convergence for the two approaches;
F12 accelerates this convergence and thus largely removes those discrepancies.
In PNO-LCCSD(T)-F12, the bond lengths are close to the experimental values with overestimation of as small as  \qtyrange{0.001}{0.003}{\angstrom}.

TB+$\Delta$MTP\#5 shows an excellent agreement to the reference PNO-LCCSD(T)-F12 with the all-electron treatment with errors about \qty{0.0003}{\angstrom} or less both for the C--C and the C--H bond lengths.
Eventually, the errors from experimental values are less than \qty{0.002}{\angstrom}.
ANI-1ccx shows a similar error for the C--C bond length but a larger error of \qty{0.004}{\angstrom} for the C--H bond length.

\subsection{Vibrational frequencies of \texorpdfstring{H\textsubscript{2}}{H2} and \texorpdfstring{C\textsubscript{6}H\textsubscript{6}}{C6H6}}
\label{sec:results:frequencies}

\begin{figure}[tb]
\centering
\includegraphics{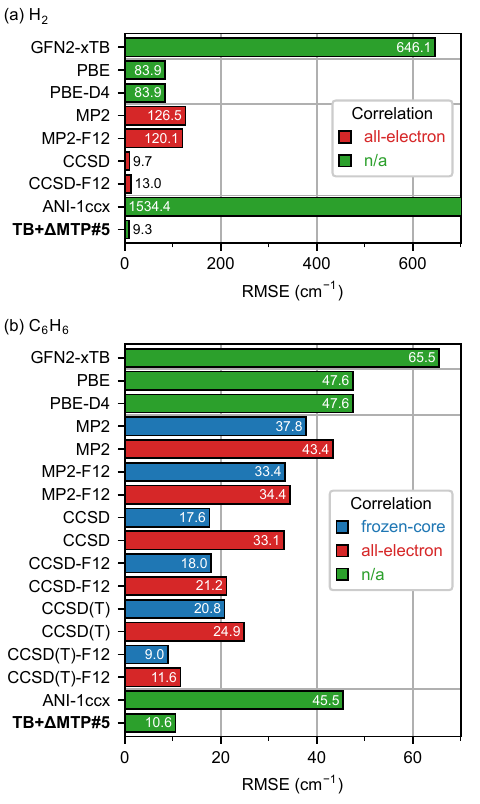}
\caption{RMSEs of harmonic vibrational frequencies with respect to the experimental values.
(a) \ce{H2}. (b) \ce{C6H6}. The heavy-aug-cc-pVTZ basis set was used for quantum-chemical calculations.}
\label{fig:vibrations}
\end{figure}

\begin{figure*}[!tb]
\centering
\includegraphics{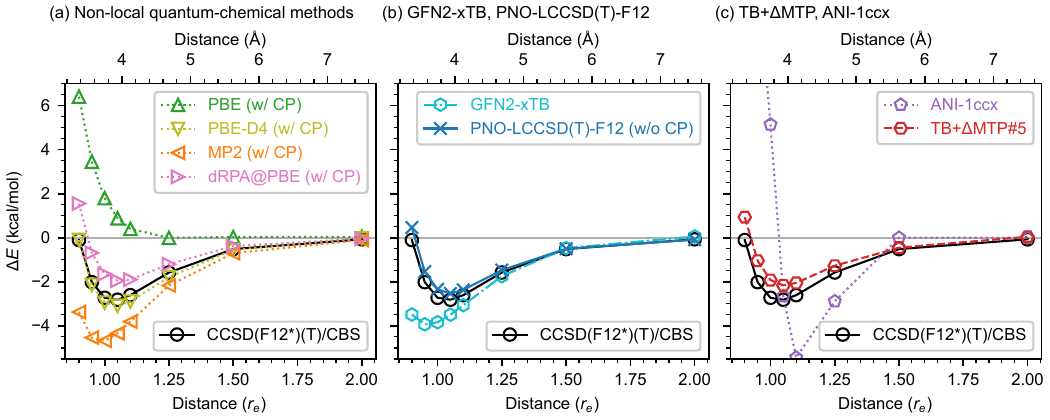}
\caption{Intermolecular interaction energies of a benzene--benzene dimer with $\pi$--$\pi$ stacking.
The geometries refer to the S66x8 dataset~\cite{Rezac_JCTC_2011_S66,*Rezac_JCTC_2014_Erratum,Brauer_PCCP_2016_S66x8}.
The solid black curve shows the CCSD(F12*)(T)/CBS values from the S66x8 dataset~\cite{Brauer_PCCP_2016_S66x8} for reference.
The $x$-axis shows the intermolecular distance scaled by the equilibrium value $r_e$ provided in the S66x8 dataset.}
\label{fig:energies_vs_MTP}
\end{figure*}

\cref{fig:vibrations} presents the RMSEs of harmonic vibrational frequencies computed with GFN2-xTB, various quantum-chemical methods, ANI-1ccx, and TB+$\Delta$MTP\#5 with respect to experimental values~\cite{Huber_Book_1979_Molecular,Handy_CPL_197_506_1992,Goodman_JPC_95_9044_1991}.
Sec.~\textcolor{violet}{S5} in the SM provides the full values of the obtained harmonic vibrational frequencies.
The harmonic vibrational frequencies were computed with the finite-displacement method with a displacement of \qty{0.01}{\angstrom} and analytical forces using our own implementation.
Note that the PNO methods currently have no implementation of analytical forces in MOLPRO and are therefore not considered in this part.

For \ce{H2} (\cref{fig:vibrations}(a)), GFN2-xTB shows a large deviation of \qty{646.1}{cm^{-1}}.
The error is largely reduced in DFT, particularly in PBE(-D4) (\qty{83.9}{cm^{-1}}).
While MP2 and MP2-F12 show larger errors than PBE (\qtylist{126.5;120.1}{cm^{-1}}, respectively), the errors in CCSD and CCSD-F12 are as small as \qtylist{9.7;13.0}{cm^{-1}}, respectively.
TB+$\Delta$MTP\#5 also shows a small RMSE comparable with CCSD and CCSD-F12, \qty{9.3}{cm^{-1}}.
In contrast, ANI-1ccx shows an error as large as \qty{1534.4}{cm^{-1}}, which is again due to the absence of \ce{H2} in the ANI-1ccx training dataset.

For \ce{C6H6} (\cref{fig:vibrations}(b)), GFN2-xTB shows an error of \qty{65.5}{cm^{-1}}, and DFT shows slightly smaller deviations (\qtyrange{47.6}{54.3}{cm^{-1}}).
For the post-HF methods, the all-electron treatment with the present heavy-aug-cc-pVTZ basis set leads to larger deviations from experiments than the frozen-core approximation.
This impact is insignificant for the F12 methods but substantial for the non-F12 methods.
CCSD(T)-F12 with the frozen-core approximation and the all-electron treatment shows errors of \qtylist{9.0;11.6}{cm^{-1}}, respectively.
The error of TB+$\Delta$MTP\#5 is \qty{10.6}{cm^{-1}}, comparable with CCSD(T)-F12.
In contrast, the error of ANI-1ccx, \qty{45.5}{cm^{-1}}, is only comparable with the DFT and the MP2 methods.
\textcolor{rev0}{The success of TB+$\Delta$MTP\#5 in reproducing the vibrational frequencies is likely due to the incorporation of MD snapshots into the training dataset.}

Thus, we may conclude that TB+$\Delta$MTP\#5 has an accuracy beyond DFT, nearly reaching the CCSD(T) level.
Note that the D4 vdW correction on PBE does not improve the vibrational frequencies, like for bond lengths (\cref{sec:results:bond_lengths}), demonstrating again that such a semi-empirical vdW-specific correction cannot fully replace quantum-chemical calculations.

\subsection{Inter-molecular interaction energies of \texorpdfstring{C\textsubscript{6}H\textsubscript{6}}{C6H6}}
\label{sec:results:dimer}

\cref{fig:energies_vs_MTP} depicts the inter-molecular interaction energy curves of a benzene--benzene dimer with $\pi$--$\pi$ stacking computed with GFN2-xTB, various quantum-chemical methods, ANI-1ccx, and TB+$\Delta$MTP\#5.
The geometries are taken from the S66x8 dataset~\cite{Rezac_JCTC_2011_S66,*Rezac_JCTC_2014_Erratum} revised in Ref.~\cite{Brauer_PCCP_2016_S66x8} alongside the reference values of the CCSD(F12*)(T) method~\cite{Haettig_JCP_2010_Communications} at the CBS limit.
For the quantum-chemical calculations, except PNO-LCCSD(T)-F12, the counterpoise (CP) correction is applied.
Below we first review the results of the quantum-chemical calculations and then discuss the quality of TB+$\Delta$MTP\#5.

In \cref{fig:energies_vs_MTP}(a), PBE cannot predict the dimer attraction even qualitatively, as well known due to the absence of the vdW interaction in this exchange--correlation functional and consistent with a previous study~\cite{Barone_JCC_30_934_2008}.
The agreement is largely improved in PBE-D4, as however the parameters in the D4 formalism are determined to reproduce the S66x8 values semi-empirically~\cite{Caldeweyher_JCP_2017_Extension,Caldeweyher_JCP_2019_generally}.
The dimer attraction is qualitatively reproduced in the MP2 method.
However, the strength of the binding is largely overestimated, and the equilibrium dimer distance is substantially underestimated.
The direct RPA (dRPA) method shows a much better agreement with the reference values compared to DFT and MP2.
However, this agreement largely owes to the CP correction.
As detailed in Sec.~\textcolor{violet}{S3} in the SM, without the CP correction, the magnitude of the binding energy is overestimated, and particularly when an all-electron treatment is employed, the deviation is significant, meaning that the CP correction is essential if the RPA methods are used to generate a training dataset for an MLIP.
However, routine applications of the CP correction are highly cumbersome, particularly for systems with more than two molecules.
In this regard, the RPA methods are not suitable for making an MLIP applicable particularly for systems with inter-molecular interactions.

In \cref{fig:energies_vs_MTP}(b), PNO-LCCSD(T)-F12 agrees well with the reference values~\cite{Brauer_PCCP_2016_S66x8} of CCSD(F12*)(T)/CBS with errors less than~\qty{0.6}{kcal/mol} across all distances examined.
Further, unlike the RPA methods above, PNO-LCCSD(T)-F12 does not reveal a substantial BSSE (Sec.~\textcolor{violet}{S1} in the SM) and thus does not require the CP correction.
Hence, the calculations of multi-molecule systems for the training of an MLIP can be performed on an equal footing as for multimers.
Note that a recent study by Hansen \textit{et al.}~\cite{Hansen_JPCA_129_4812_2025} demonstrates that further refining of the domain and the pair approximations in PNO-LCCSD(T)-F12 increases the magnitude of the equilibrium interaction energy by about \qty{0.1}{kcal/mol}, thereby narrowing the gap with the reference CCSD(T)/CBS value, though at the expense of longer computation times.
The agreement of GFN2-xTB with the reference CCSD(F12*)(T)/CBS values~\cite{Brauer_PCCP_2016_S66x8} is better than MP2.
Since the computational cost of GFN2-xTB is much cheaper than DFT and post-HF, GFN2-xTB is a better baseline for the $\Delta$-learning approach.

In \cref{fig:energies_vs_MTP}(c), TB+$\Delta$MTP\#5 well reproduces the PNO-LCCSD(T)-F12 values and thus also the reference S66x8 values~\cite{Brauer_PCCP_2016_S66x8}, with an error of less than \qty{0.6}{kcal/mol} at $r_\mathrm{e} = 1.00$.
In contrast, ANI-1ccx cannot reproduce the reference energies even qualitatively.
This is likely due to a bias in the ANI-1ccx dataset.
Specifically, while the ANI-1ccx dataset~\cite{Smith_NC_2019_Approaching,Smith_SD_2020_ANI} includes as many as \num{500000} structures, most of them are monomers of relatively small molecules and thus do not include London dispersion interaction.
Furthermore, the interaction energies predicted by ANI-1cxx fall off to zero for distances larger than $1.5 r_e \approx \qty{5.6}{\angstrom}$ (larger than the cutoff radius of ANI-1cxx) due to the absence of any long-range interactions.

\subsection{Application to the \texorpdfstring{C\textsubscript{48}H\textsubscript{30}}{C48H30} COF\label{sec:results:C48H30}}

\begin{figure}[tb]
\centering
\includegraphics{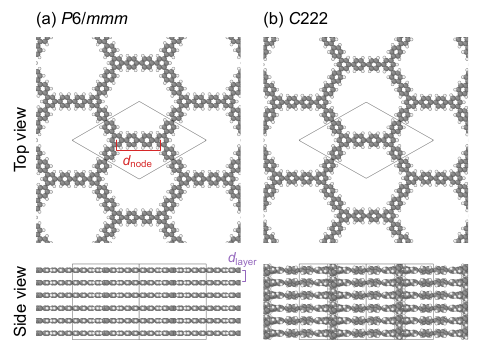}
\caption{Structures of the \ce{C48H30} COF optimized with TB+$\Delta$MTP\#5.}
\label{fig:C48H30}
\end{figure}

\begin{figure}[tb]
\centering
\includegraphics{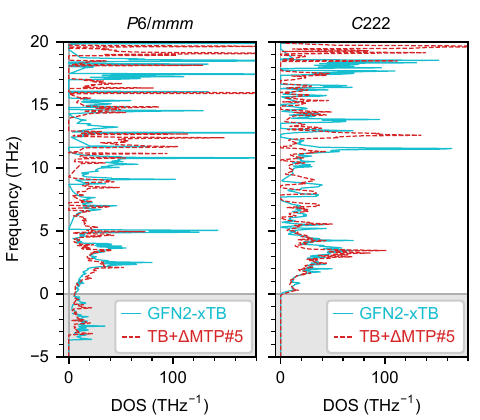}
\caption{Phonon DOSs of the \ce{C48H30} COF obtained with GFN2-xTB and TB+$\Delta$MTP\#5.
The negative-frequency region corresponds to imaginary modes.}
\label{fig:C48H30:phonons}
\end{figure}

Having thus verified the performance of TB+$\Delta$MTP\#5, we next apply this potential to the periodic \ce{C48H30} COF.
For comparison, the results with GFN2-xTB and PBE-D4 are also shown.
All calculations employed a six-layer simulation cell containing 468 atoms.
For the tight-binding and the DFT calculations, the reciprocal spaces of the \textcolor{rev0}{468-atom} simulation cells were sampled by the $\Gamma$-point-only mesh.
Further details on the DFT calculations are provided in Sec.~\textcolor{violet}{S6} in the SM.

We first considered fully eclipsed layer stacking with perfectly planar layers, as shown in \cref{fig:C48H30}(a), which shows a space group type of $P6/mmm$ (No.\,191).
However, its phonon density of states (DOS) (\cref{fig:C48H30:phonons}(a)) shows numerous imaginary modes, indicating that this structure is dynamically unstable at \qty{0}{K}.
Therefore, to further relax the structure, we once broke its symmetry by perturbing the positions of the atoms by a Gaussian distribution with a standard variation of \qty{0.01}{\angstrom} and then re-relaxed the structure.
After the relaxation and symmetry refinement, the structure shows a space group type of $C222$ (No.\,21).
In this structure, the layers are not fully planar anymore but rather show a twisting between in-plane adjacent benzene rings, as shown in \cref{fig:C48H30}(b).
The energy of the $C222$ structure is lower by \qty{10.2}{meV/atom} than the $P6/mmm$ structure, as given in \cref{tab:C48H30:energies}.
The phonon DOS of the $C222$ structure shows no imaginary modes (\cref{fig:C48H30:phonons}(b)), supporting its dynamical stability.
The same trend is obtained also for GFN2-xTB and the PBE-D4 vdW DFT functional.
The twisting is well consistent with molecular systems like biphenyl~\cite{Almenningen_JMS_128_59_1985}.

\cref{tab:C48H30:energies} also summarizes the obtained node and inter-layer distances of the \ce{C48H30} COF.
The node-distances, i.e., the distances between the centers of the benzene rings at the end of the $p$-quaterphenyl part, of the $C222$ \ce{C48H30} COF are \qty{12.891}{\angstrom} and in good agreement with experimentally measured values of \qty{12.9(6)}{\angstrom}~\cite{Blunt_CC_46_7157_2010} and \qty{13.6(6)}{\angstrom}~\cite{Shi_CC_52_8726_2016}.
The inter-layer distances of the $C222$ \ce{C48H30} COF are \qty{3.673}{\angstrom}, substantially larger than those of graphite in experiments, \qtyrange{3.35}{3.36}{\angstrom}~\cite{Trucano_N_1975_Structure,Zhao_PRB_1989_X}.
The inter-layer distances with TB+$\Delta$MTP\#5 are substantially larger than the values in GFN2-xTB while in a similar range to PBE-D4, consistent with the inter-molecular distances for a benzene--benzene dimer in \cref{fig:energies_vs_MTP}.

\begin{table}[!tb]
\small
\centering
\caption{Relative energies $\Delta E$ (meV/atom) of the $C222$ structure with respect to $P6/mmm$ and inter-layer distances $d_\mathrm{layer}$~(\unit{\angstrom}) of the \ce{C48H30} COF.}
\label{tab:C48H30:energies}
\begin{tabular}{lS[table-format=4.1,table-auto-round]*{4}{S[table-format=3.3,table-auto-round]}}
\toprule
&
&
\multicolumn{2}{c}{$d_\mathrm{node}$} &
\multicolumn{2}{c}{$d_\mathrm{layer}$} \\
&
\multicolumn{1}{c}{$\Delta E$} &
\multicolumn{1}{c}{$P6/mmm$} &
\multicolumn{1}{c}{$C222$} &
\multicolumn{1}{c}{$P6/mmm$} &
\multicolumn{1}{c}{$C222$} \\
\midrule
GFN2-xTB       & -11.8      & 12.961948784 & 12.766446150 & 3.325034753 & 3.427297986 \\
PBE-D4         & -10.7      & 13.031535908 & 12.847782666 & 3.667272789 & 3.677571281 \\
\bfTBDMTP[5]   & -10.182870 & 13.007031067 & 12.891463546 & 3.683139 & 3.672534 \\
\bottomrule
\end{tabular}
\end{table}

\begin{figure*}[!tb]
    \centering
    \includegraphics{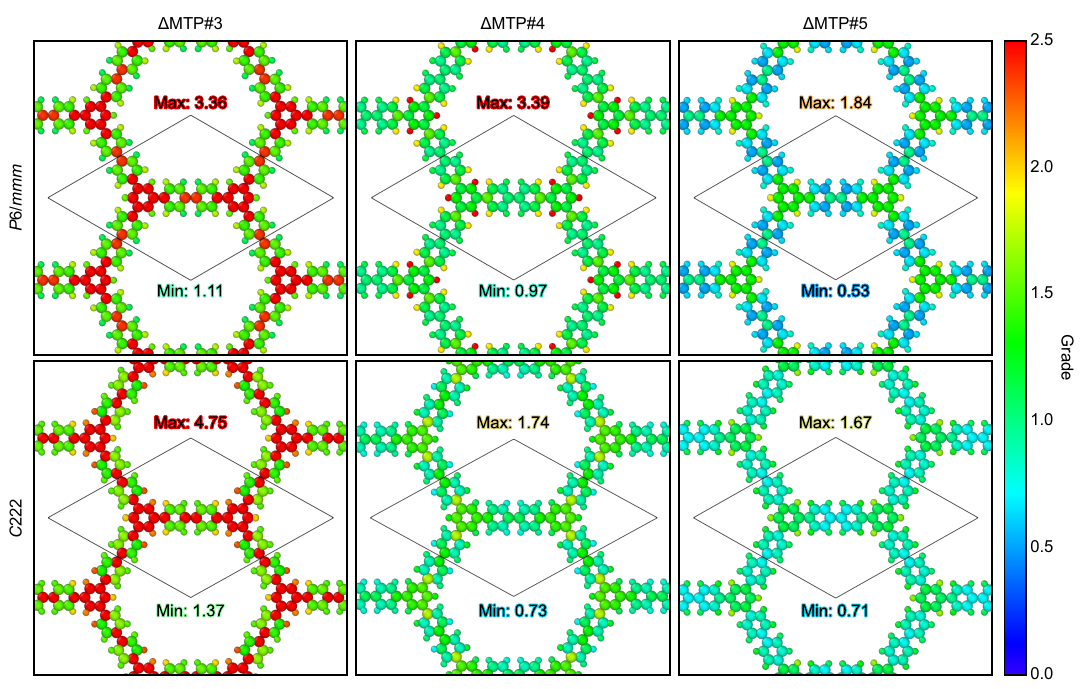}
    \caption{Local extrapolation grades of the \ce{C48H30} COF obtained with $\Delta$MTPs.}
    \label{fig:nbh_grades}
\end{figure*}

Based on the findings of layer offsets in previous atomistic simulations and experiments for general quasi-2D COFs \cite{
Lukose_CEJ_17_2388_2011,  Spitler_JACS_133_19416_2011,  Kang_JACS_142_12995_2020,  Puetz_CS_11_12647_2020,  Emmerling_JACS_143_15711_2021,  Kessler_MMM_336_111796_2022},  we also investigated the shearing from the $C222$ structure, but the obtained structure was substantially higher in energy then the unsheared one.
Notably, the detailed symmetry of COFs was not discussed in the previous database~\cite{Ongari_ACS_2019_Building}.
The results imply the possibility for further improvement of COF structures in the existing database, which may eventually improve the prediction of COF properties based on such a database.

\cref{fig:nbh_grades} shows the local extrapolation grades obtained with $\Delta$MTPs for the \ce{C48H30} COF with the two structures.
Like the molecular test dataset (\cref{fig:validation}(c)), the extrapolation grades decrease with increasing training-dataset size from \#3 toward \#5.
Particularly for $\Delta$MTP\#5, the grades are less than 2 for all the atoms, indicating that the prediction for the present COF should be reliable.
It is worth emphasizing that, although the direct energy comparison with the reference CCSD(T)-level energy for the periodic COF is impractical due to the computational cost, we can still evaluate the transferability of the $\Delta$MTPs with the extrapolation grades as demonstrated.

\begin{figure}[!tb]
    \centering
    \includegraphics{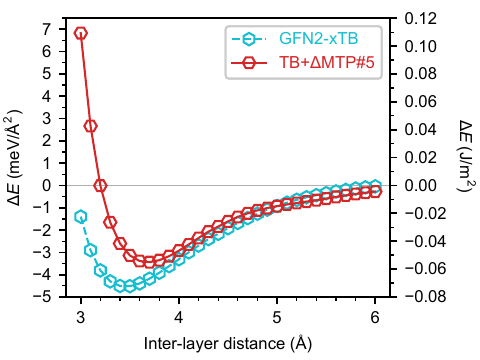}
    \caption{Inter-layer binding energies of the \ce{C48H30} COF in the $C222$ symmetry computed with GFN2-xTB and TB+$\Delta$MTP\#5 as a function of inter-layer distance.}
    \label{fig:C48H30:interlayer_binding_energies}
\end{figure}

\cref{fig:C48H30:interlayer_binding_energies} shows the inter-layer binding energies of the \ce{C48H30} COF with the $C222$ symmetry computed with GFN2-xTB and TB+$\Delta$MTP\#5 as a function of inter-layer distance.
The calculations are performed by modifying the inter-layer distance and fixing the intra-layer geometry.
The reference single-layer energy is obtained using cells with vacuum layer thicknesses larger than \qty{40}{\angstrom}.
TB+$\Delta$MTP\#5 shows a longer equilibrium inter-layer distance and lower inter-layer binding energies in magnitude than GFN2-xTB, consistent with our findings for the benzene--benzene dimer (\cref{fig:energies_vs_MTP}).
The binding energy in TB+$\Delta$MTP\#5 is \qty{0.055}{J/m\textsuperscript{2}}.
This is nearly four times smaller than the inter-layer binding energy of graphite in experiments, \qtyrange{0.19}{0.37}{J/m\textsuperscript{2}}~\cite{Girifalco_JCP_25_693_1956,Benedict_CPL_286_490_1998,Zacharia_PRB_2004_Interlayer,Liu_PRB_2012_Interlayer}, which is qualitatively reasonable because the \ce{C48H30} COF has a more sparse in-layer atomic density than graphite and thus shows weaker London dispersion interaction.

\begin{figure}[tbp!]
    \centering
    \includegraphics{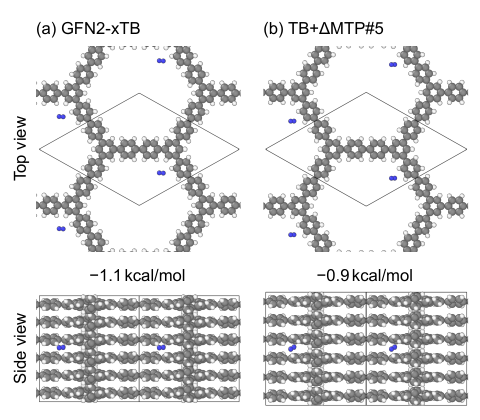}
    \caption{Absorption of an \ce{H2} molecule (blue) in the \ce{C48H30} COF with the $C222$ symmetry obtained using (a)~GFN2-xTB and (b)~TB+$\Delta$MTP\#5.
    The corresponding absorption energies are also shown.}
    \label{fig:C48H30+H2}
\end{figure}

We finally investigate hydrogen absorption on the \ce{C48H30} COF.
To compute the \ce{H2} absorption energies, we put an \ce{H2} molecule with a random orientation at a random position in the simulation cell and then optimize the structure, while fixing the COF lattice following a previous study~\cite{Han_JACS_2008_Covalent}.
We do 10 such calculations for each of GFN2-xTB and TB+$\Delta$MTP\#5.
\cref{fig:C48H30+H2} presents the optimized \ce{H2} absorption sites with the most negative absorption energies.
In both methods, the \ce{H2} absorption sites with the most negative absorption energies are found near the node position between two adjacent layers.
The absorption is weaker in TB+$\Delta$MTP\#5 (\qty{-0.9}{kcal/mol}), thus effectively in CCSD(T), than for GFN2-xTB (\qty{-1.1}{kcal/mol}).
It is here worth re-emphasizing that, while TB+$\Delta$MTP\#5 exhibits near CCSD(T) accuracy, its evaluation is computationally even less expensive than DFT and that many such evaluations are affordable. Together with the prediction of phonon DOS, this will be very useful for quantitative predictions in COF research.

\section{Conclusions}
\label{sec:conclusions}

\color{rev0}
We have developed a methodology to train MLIPs with
the gold-standard accuracy of quantum chemistry, CCSD(T), particularly suitable for systems with extended covalent networks and
long‑range vdW interactions.
Based on the $\Delta$-learning approach with a tight-binding baseline, the MLIP can be trained solely with molecular systems,
including dispersion‑dominated configurations.
The thus trained MLIP delivers an RMSE as low as \qty{0.4}{meV/atom} with respect to the reference PNO-CCSD(T)-F12 values and is also seamlessly transferable from molecular to crystalline systems.
\color{black}
We have showcased its capabilities by resolving the detailed structural symmetry, the harmonic vibrational frequencies, the inter-layer binding energy, and the hydrogen absorption energy of the prototypical COF, \ce{C48H30}.

Although the present TB+$\Delta$MTP model is tailored to the \ce{C48H30} COF, the workflow itself is transferable to other systems.
Expanding the training set to chemically diverse fragments will broaden the applicability of the potential.
The $\Delta$-learning strategy is also straightforward for porting to alternative MLIP architectures.
Such extensions will empower high‑throughput, CCSD(T)‑quality screening of large libraries of COFs and other vdW‑dominated materials, thereby accelerating materials discovery.

\section*{Funding}  

YI, JJ, and BG acknowledge the funding by the Deutsche Forschungsgemeinschaft (DFG, German Research Foundation) under SFB 1333/2 2022 (Project No.~358283783).
YI, YO, AK, and BG acknowledge the funding by the DFG under Germany's Excellence Strategy --- EXC 2075 (Project No.~390740016) with the support by the Stuttgart Center for Simulation Science (SC SimTech).
YI and PK acknowledge the funding by the DFG under IK 125/1-1 (Project No.~519607530).
YI, AF, JJ, and BG acknowledge the funding from the European Research Council (ERC) under the European Union’s Horizon 2020 research and innovation programme (grant agreement No. 865855).
YI and AF acknowledge the support by the Faculty of Chemistry of the University of Stuttgart.
This project is also supported by the DFG under INST 40/575-1 FUGG --- JUSTUS 2 cluster (Project No.~405998092).

\section*{Data availability}

All PNO-LCCSD(T)-F12 data used for the MLIP training is available at \url{https://darus.uni-stuttgart.de/previewurl.xhtml?token=eb641dad-d8cd-489a-b5a7-43192b3276b7} (temporary link for review).

\section*{Code availability}

Our MTP implementation is available at \url{https://github.com/imw-md/motep}.

\section*{Author contributions}

YI contributed to conceptualization and methodology development, carried out the investigation, and drafted the original manuscript. All authors contributed to the analysis and interpretation of the results and revised the manuscript. YI, AF, and PK developed our implementation of \textcolor{rev0}{MTPs}. YO and JJ assisted with the use of software to compute local extrapolation grades. AK and BG contributed to project administration.

\section*{Competing interests}

The authors declare no competing interests.

\end{document}


\title{Supplemental Materials:\texorpdfstring{\\}{ }Machine-learning interatomic potentials achieving CCSD(T) accuracy for systems with extended covalent networks and van der Waals interactions}

\author{Yuji Ikeda\,\orcidlink{0000-0001-9176-3270}}
\email{yuji.ikeda@imw.uni-stuttgart.de}
\affiliation{Institute for Materials Science,
\href{https://ror.org/04vnq7t77}{University of Stuttgart},
Pfaffenwaldring 55, 70569 Stuttgart, Germany}

\author{Axel Forslund\,\orcidlink{0000-0002-0419-3546}}
\affiliation{Institute for Materials Science,
\href{https://ror.org/04vnq7t77}{University of Stuttgart},
Pfaffenwaldring 55, 70569 Stuttgart, Germany}
\affiliation{Department of Materials Science and Engineering,
\href{https://ror.org/026vcq606}{KTH Royal Institute of Technology},
SE-100 44 Stockholm, Sweden}

\author{Pranav Kumar\,\orcidlink{0000-0002-3661-5870}}
\affiliation{Institute for Materials Science,
\href{https://ror.org/04vnq7t77}{University of Stuttgart},
Pfaffenwaldring 55, 70569 Stuttgart, Germany}

\author{Yongliang~Ou\,\orcidlink{0000-0003-1486-1197}}
\affiliation{Institute for Materials Science,
\href{https://ror.org/04vnq7t77}{University of Stuttgart},
Pfaffenwaldring 55, 70569 Stuttgart, Germany}
\affiliation{Department of Materials Science and Engineering,
\href{https://ror.org/042nb2s44}{Massachusetts Institute of Technology},
77 Massachusetts Avenue, Cambridge, 02139, MA, USA}

\author{Jong Hyun Jung\,\orcidlink{0000-0002-2409-975X}}
\affiliation{Institute for Materials Science,
\href{https://ror.org/04vnq7t77}{University of Stuttgart},
Pfaffenwaldring 55, 70569 Stuttgart, Germany}

\author{Andreas Köhn\,\orcidlink{0000-0002-0844-842X}}
\affiliation{Institute for Theoretical Chemistry,
\href{https://ror.org/04vnq7t77}{University of Stuttgart},
Pfaffenwaldring 55, 70569 Stuttgart, Germany}

\author{Blazej Grabowski\,\orcidlink{0000-0003-4281-5665}}
\affiliation{Institute for Materials Science,
\href{https://ror.org/04vnq7t77}{University of Stuttgart},
Pfaffenwaldring 55, 70569 Stuttgart, Germany} 
\maketitle

\section{Basis-set-superposition error (BSSE)}
\label{sec:BSSE}

Intermolecular dispersion interaction energies obtained in quantum-chemical calculations are in general influenced by the basis set superposition error (BSSE); the stability of a multimer is overestimated by the monomer energies evaluated with the basis sets of the whole multimer.
It is therefore in principle necessary to compensate this spurious effect in order to evaluate the intermolecular interaction energies.
The counterpoise (CP) correction by Boys and Bernardi~\cite{Boys_MP_1970_calculation} is one of the common methods for this compensation.

\begin{figure*}[!tb]
\centering
\includegraphics{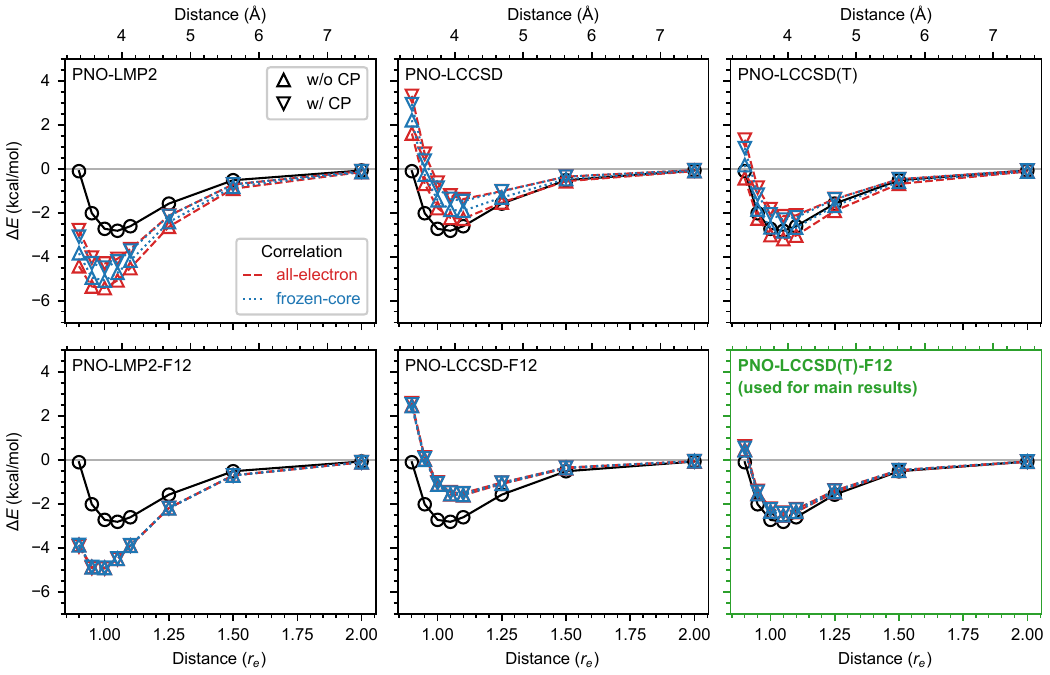}
\caption{Inter-molecular interaction energies of a benzene--benzene dimer with $\pi$--$\pi$ stacking obtained with various local correlation methods (kcal/mol).
The reference CCSD(F12*)(T)/CBS values~\cite{Brauer_PCCP_2016_S66x8} are also shown for comparison in the solid black curves.}
\label{fig:counterpoise}
\end{figure*}

\cref{fig:counterpoise} and \cref{tab:counterpoise} present the inter-molecular interaction energies of a benzene--benzene dimer with $\pi$--$\pi$ stacking in the S66x8 dataset~\cite{Rezac_JCTC_2011_S66,*Rezac_JCTC_2014_Erratum} obtained with various local correlation methods without and with the CP correction.
The heavy-aug-cc-pVTZ basis set is employed.
In the calculations with core-electron excitation, the excitation is considered for both the correlation part and the CABS singles correction.

\begin{table}[tb]
\centering
\caption{Inter-molecular interaction energies of a benzene--benzene dimer with $\pi$--$\pi$ stacking obtained with various local correlation methods~(kcal/mol).}
\label{tab:counterpoise}
\begin{tabular}{lcc*{8}{S[table-format=4.4,retain-explicit-plus]}}
\toprule
& & & \multicolumn{8}{c}{$r_e$} \\
\cmidrule{4-11}
\multicolumn{1}{c}{Method} &
\multicolumn{1}{c}{Correlation} &
\multicolumn{1}{c}{CP} &
\multicolumn{1}{c}{0.90} &
\multicolumn{1}{c}{0.95} &
\multicolumn{1}{c}{1.00} &
\multicolumn{1}{c}{1.05} &
\multicolumn{1}{c}{1.10} &
\multicolumn{1}{c}{1.25} &
\multicolumn{1}{c}{1.50} &
\multicolumn{1}{c}{2.00} \\
\midrule
        PNO-LMP2 &  frozen-core & w/o CP & -3.835 & -4.942 & -5.074 & -4.728 & -4.170 & -2.420 & -0.811 & -0.142 \\
        PNO-LMP2 &  frozen-core &  w/ CP & -3.083 & -4.288 & -4.492 & -4.210 & -3.715 & -2.137 & -0.702 & -0.104 \\
        PNO-LMP2 & all-electron & w/o CP & -4.424 & -5.348 & -5.417 & -5.067 & -4.510 & -2.630 & -0.901 & -0.143 \\
        PNO-LMP2 & all-electron &  w/ CP & -2.783 & -4.031 & -4.302 & -4.088 & -3.646 & -2.129 & -0.706 & -0.104 \\
    PNO-LMP2-F12 &  frozen-core & w/o CP & -3.885 & -4.878 & -4.912 & -4.498 & -3.910 & -2.201 & -0.710 & -0.104 \\
    PNO-LMP2-F12 &  frozen-core &  w/ CP & -3.866 & -4.863 & -4.895 & -4.474 & -3.878 & -2.181 & -0.700 & -0.102 \\
    PNO-LMP2-F12 & all-electron & w/o CP & -3.863 & -4.867 & -4.907 & -4.499 & -3.900 & -2.191 & -0.705 & -0.108 \\
    PNO-LMP2-F12 & all-electron &  w/ CP & -3.930 & -4.916 & -4.935 & -4.506 & -3.894 & -2.186 & -0.702 & -0.102 \\
       PNO-LCCSD &  frozen-core & w/o CP & +2.212 & -0.263 & -1.426 & -1.866 & -1.904 & -1.302 & -0.463 & -0.101 \\
       PNO-LCCSD &  frozen-core &  w/ CP & +2.947 & +0.377 & -0.858 & -1.359 & -1.458 & -1.021 & -0.349 & -0.057 \\
       PNO-LCCSD & all-electron & w/o CP & +1.609 & -0.679 & -1.782 & -2.225 & -2.267 & -1.528 & -0.561 & -0.101 \\
       PNO-LCCSD & all-electron &  w/ CP & +3.321 & +0.689 & -0.627 & -1.211 & -1.371 & -1.008 & -0.352 & -0.056 \\
   PNO-LCCSD-F12 &  frozen-core & w/o CP & +2.447 & +0.007 & -1.123 & -1.554 & -1.594 & -1.081 & -0.369 & -0.063 \\
   PNO-LCCSD-F12 &  frozen-core &  w/ CP & +2.523 & +0.073 & -1.060 & -1.487 & -1.526 & -1.037 & -0.347 & -0.055 \\
   PNO-LCCSD-F12 & all-electron & w/o CP & +2.481 & +0.026 & -1.118 & -1.568 & -1.606 & -1.092 & -0.378 & -0.070 \\
   PNO-LCCSD-F12 & all-electron &  w/ CP & +2.585 & +0.117 & -1.027 & -1.470 & -1.506 & -1.029 & -0.347 & -0.055 \\
    PNO-LCCSD(T) &  frozen-core & w/o CP & +0.172 & -1.845 & -2.662 & -2.840 & -2.671 & -1.680 & -0.585 & -0.118 \\
    PNO-LCCSD(T) &  frozen-core &  w/ CP & +0.944 & -1.174 & -2.066 & -2.307 & -2.203 & -1.385 & -0.465 & -0.072 \\
    PNO-LCCSD(T) & all-electron & w/o CP & -0.425 & -2.255 & -3.013 & -3.193 & -3.031 & -1.906 & -0.683 & -0.118 \\
    PNO-LCCSD(T) & all-electron &  w/ CP & +1.328 & -0.852 & -1.827 & -2.152 & -2.110 & -1.369 & -0.468 & -0.071 \\
PNO-LCCSD(T)-F12 &  frozen-core & w/o CP & +0.436 & -1.554 & -2.344 & -2.516 & -2.354 & -1.457 & -0.489 & -0.080 \\
PNO-LCCSD(T)-F12 &  frozen-core &  w/ CP & +0.549 & -1.454 & -2.250 & -2.422 & -2.262 & -1.399 & -0.462 & -0.070 \\
PNO-LCCSD(T)-F12 & all-electron & w/o CP & +0.465 & -1.538 & -2.341 & -2.532 & -2.367 & -1.469 & -0.499 & -0.087 \\
PNO-LCCSD(T)-F12 & all-electron &  w/ CP & +0.612 & -1.408 & -2.214 & -2.403 & -2.241 & -1.390 & -0.463 & -0.070 \\
\midrule
CCSD(F12*)T/CBS\footnotemark[1]
                 &  frozen-core &  w/ CP & -0.105 & -2.016 & -2.725 & -2.813 & -2.607 & -1.578 & -0.515 & -0.072 \\
\bottomrule
\end{tabular}
 \footnotetext[1]{Brauer~\textit{et al.}~\cite{Brauer_PCCP_2016_S66x8}}
\end{table}

Without the F12 explicit electronic correlation (top row), the impact of the CP correction is substantial and larger in the all-electron treatment (\qtyrange{1.1}{1.2}{kcal/mol}) than in the frozen-core approximation (\qty{0.6}{kcal/mol}) at $r_e = 1.00$.
In contrast, with the F12 explicit correlation (bottom row), the impact of the CP correction is marginal and less than \qty{0.1}{kcal/mol}.
Thus, the combination of the local correlation methods and the F12 explicit correlation method significantly reduces the BSSE, consistent with the report by Ma and Werner~\cite{Ma_WCMS_2018_Explicitly}.
The BSSE in these approaches is much smaller than the accuracy that can be achieved by the present MLIP formalism, i.e., \qtyrange{0.1}{1}{meV/atom} and hence essentially negligible for the training of MLIPs. 
Further, the core-electron treatment does not substantially affect the inter-molecular interaction energies, unlike TAEs (Sec.~III\,A in the main text).

 \section{Basis-set dependence}
\label{sec:basis_set_dependence}

\begin{figure*}
\centering
\includegraphics{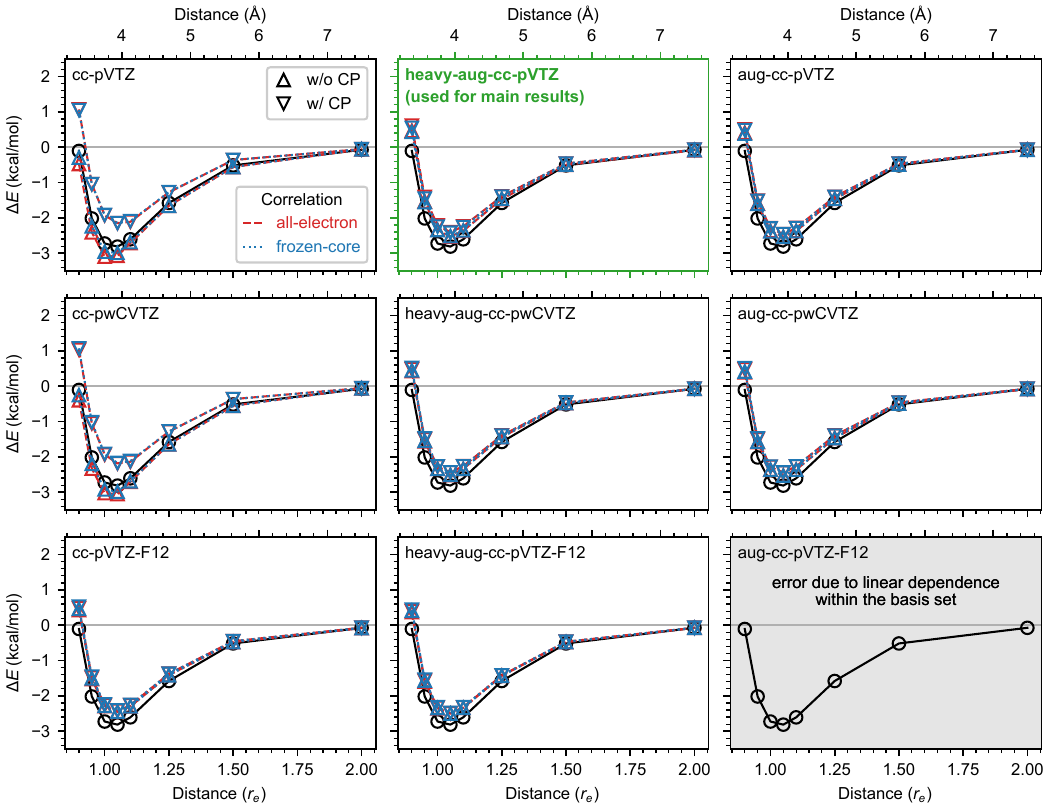}
\caption{Basis-set dependence of the inter-molecular interaction energies of a benzene--benzene dimer with $\pi$--$\pi$ stacking obtained with PNO-LCCSD(T)-F12 (kcal/mol).
The reference CCSD(F12*)(T)/CBS values~\cite{Brauer_PCCP_2016_S66x8} are also shown for comparison in the solid black curves.}
\label{fig:basis_set_dependence}
\end{figure*}

\begin{table}
\centering
\caption{Basis-set dependence of the inter-molecular interaction energies of a benzene--benzene dimer with $\pi$--$\pi$ stacking obtained with PNO-LCCSD(T)-F12 (kcal/mol).}
\label{tab:basis_set_dependence}
\begin{tabular}{lcc*{8}{S[table-format=4.4,retain-explicit-plus]}}
\toprule
& & & \multicolumn{8}{c}{$r_e$} \\
\cmidrule{4-11}
\multicolumn{1}{c}{Basis set} &
\multicolumn{1}{c}{Correlation} &
\multicolumn{1}{c}{CP} &
\multicolumn{1}{c}{0.90} &
\multicolumn{1}{c}{0.95} &
\multicolumn{1}{c}{1.00} &
\multicolumn{1}{c}{1.05} &
\multicolumn{1}{c}{1.10} &
\multicolumn{1}{c}{1.25} &
\multicolumn{1}{c}{1.50} &
\multicolumn{1}{c}{2.00} \\
\midrule
                 cc-pVTZ &  frozen-core & w/o CP & -0.294 & -2.241 & -2.955 & -2.992 & -2.685 & -1.659 & -0.561 & -0.059 \\
                 cc-pVTZ &  frozen-core &  w/ CP & +1.051 & -1.048 & -1.919 & -2.146 & -2.092 & -1.271 & -0.361 & -0.054 \\
                 cc-pVTZ & all-electron & w/o CP & -0.475 & -2.409 & -3.098 & -3.078 & -2.714 & -1.660 & -0.561 & -0.059 \\
                 cc-pVTZ & all-electron &  w/ CP & +1.071 & -1.044 & -1.928 & -2.149 & -2.111 & -1.277 & -0.363 & -0.054 \\
       heavy-aug-cc-pVTZ &  frozen-core & w/o CP & +0.436 & -1.554 & -2.344 & -2.516 & -2.354 & -1.457 & -0.489 & -0.080 \\
       heavy-aug-cc-pVTZ &  frozen-core &  w/ CP & +0.549 & -1.454 & -2.250 & -2.422 & -2.262 & -1.399 & -0.462 & -0.070 \\
       heavy-aug-cc-pVTZ & all-electron & w/o CP & +0.465 & -1.538 & -2.341 & -2.532 & -2.367 & -1.469 & -0.499 & -0.087 \\
       heavy-aug-cc-pVTZ & all-electron &  w/ CP & +0.612 & -1.408 & -2.214 & -2.403 & -2.241 & -1.390 & -0.463 & -0.070 \\
             aug-cc-pVTZ &  frozen-core & w/o CP & +0.383 & -1.606 & -2.382 & -2.539 & -2.375 & -1.464 & -0.489 & -0.075 \\
             aug-cc-pVTZ &  frozen-core &  w/ CP & +0.484 & -1.518 & -2.300 & -2.459 & -2.291 & -1.408 & -0.466 & -0.070 \\
             aug-cc-pVTZ & all-electron & w/o CP & +0.411 & -1.593 & -2.370 & -2.538 & -2.371 & -1.464 & -0.485 & -0.070 \\
             aug-cc-pVTZ & all-electron &  w/ CP & +0.505 & -1.516 & -2.293 & -2.459 & -2.287 & -1.404 & -0.466 & -0.070 \\
               cc-pwCVTZ &  frozen-core & w/o CP & -0.250 & -2.197 & -2.920 & -2.986 & -2.689 & -1.645 & -0.555 & -0.059 \\
               cc-pwCVTZ &  frozen-core &  w/ CP & +1.074 & -1.020 & -1.917 & -2.173 & -2.117 & -1.280 & -0.365 & -0.054 \\
               cc-pwCVTZ & all-electron & w/o CP & -0.397 & -2.330 & -3.026 & -3.053 & -2.703 & -1.643 & -0.553 & -0.058 \\
               cc-pwCVTZ & all-electron &  w/ CP & +1.032 & -1.053 & -1.935 & -2.184 & -2.137 & -1.285 & -0.364 & -0.052 \\
     heavy-aug-cc-pwCVTZ &  frozen-core & w/o CP & +0.452 & -1.546 & -2.338 & -2.509 & -2.340 & -1.439 & -0.480 & -0.075 \\
     heavy-aug-cc-pwCVTZ &  frozen-core &  w/ CP & +0.518 & -1.480 & -2.269 & -2.437 & -2.268 & -1.396 & -0.460 & -0.069 \\
     heavy-aug-cc-pwCVTZ & all-electron & w/o CP & +0.429 & -1.557 & -2.341 & -2.512 & -2.339 & -1.439 & -0.480 & -0.076 \\
     heavy-aug-cc-pwCVTZ & all-electron &  w/ CP & +0.483 & -1.502 & -2.277 & -2.439 & -2.261 & -1.393 & -0.459 & -0.069 \\
           aug-cc-pwCVTZ &  frozen-core & w/o CP & +0.417 & -1.569 & -2.362 & -2.527 & -2.362 & -1.465 & -0.490 & -0.079 \\
           aug-cc-pwCVTZ &  frozen-core &  w/ CP & +0.512 & -1.480 & -2.277 & -2.441 & -2.271 & -1.396 & -0.463 & -0.070 \\
           aug-cc-pwCVTZ & all-electron & w/o CP & +0.394 & -1.579 & -2.364 & -2.530 & -2.362 & -1.465 & -0.491 & -0.080 \\
           aug-cc-pwCVTZ & all-electron &  w/ CP & +0.477 & -1.500 & -2.284 & -2.445 & -2.267 & -1.393 & -0.462 & -0.070 \\
             cc-pVTZ-F12 &  frozen-core & w/o CP & +0.466 & -1.503 & -2.278 & -2.446 & -2.294 & -1.415 & -0.477 & -0.082 \\
             cc-pVTZ-F12 &  frozen-core &  w/ CP & +0.526 & -1.454 & -2.243 & -2.410 & -2.254 & -1.372 & -0.451 & -0.070 \\
             cc-pVTZ-F12 & all-electron & w/o CP & +0.431 & -1.527 & -2.289 & -2.452 & -2.297 & -1.412 & -0.476 & -0.082 \\
             cc-pVTZ-F12 & all-electron &  w/ CP & +0.488 & -1.481 & -2.260 & -2.424 & -2.265 & -1.374 & -0.451 & -0.070 \\
   heavy-aug-cc-pVTZ-F12 &  frozen-core & w/o CP & +0.410 & -1.567 & -2.341 & -2.501 & -2.340 & -1.433 & -0.477 & -0.073 \\
   heavy-aug-cc-pVTZ-F12 &  frozen-core &  w/ CP & +0.445 & -1.543 & -2.320 & -2.479 & -2.315 & -1.418 & -0.466 & -0.070 \\
   heavy-aug-cc-pVTZ-F12 & all-electron & w/o CP & +0.370 & -1.593 & -2.351 & -2.503 & -2.339 & -1.430 & -0.477 & -0.074 \\
   heavy-aug-cc-pVTZ-F12 & all-electron &  w/ CP & +0.404 & -1.570 & -2.336 & -2.490 & -2.324 & -1.419 & -0.467 & -0.071 \\
\midrule
CCSD(F12*)T/CBS\footnotemark[1]
                         &  frozen-core &  w/ CP & -0.105 & -2.016 & -2.725 & -2.813 & -2.607 & -1.578 & -0.515 & -0.072 \\
\bottomrule
\end{tabular}
 \footnotetext[1]{Brauer~\textit{et al.}~\cite{Brauer_PCCP_2016_S66x8}}
\end{table}

The correlation-consistent polarized valence basis sets of Dunning~\cite{Dunning_JCP_1989_Gaussian}, possibly augmented with diffuse functions~\cite{Kendall_JCP_1992_Electron}, are the standard of modern quantum-chemical calculations.
However, they were optimized specifically for the frozen-core approximation and for the non-F12 method.
Therefore, various extensions have been developed for the all-electron treatment and the F12 method.
Namely, the (aug-)cc-pwCV\textit{X}Z basis sets~\cite{Peterson_JCP_117_10548_2002,DeYonker_JPCA_111_11383_2007} account for core--core and core--valence correlation effects.
The (aug-)cc-pV\textit{X}Z-F12 basis sets~\cite{Peterson_JCP_128_084102_2008,Sylvetsky_JCP_147_134106_2017} are tuned for the F12 method.
The cc-pCV\textit{X}Z-F12 basis sets~\cite{Hill_JCP_132_054108_2010,Hill_PCCP_12_10460_2010} consider both, core--core/valence correlation effects and a combination with the F12 method.

\cref{fig:basis_set_dependence} and \cref{tab:basis_set_dependence} present the inter-molecular interaction energies of a benzene--benzene dimer with $\pi$--$\pi$ stacking in the PNO-LCCSD(T)-F12 method together with the various triple-$\zeta$ basis sets.
The ``heavy'' basis sets have the augmenting diffuse functions only for non-hydrogen atoms.
The calculations with the aug-cc-pVTZ-F12 and cc-pwCVTZ-F12 basis sets failed due to intrinsic linear dependence within the basis sets.

The impact of the all-electron treatment is almost negligible irrespective of the investigated basis sets, including those tuned for the calculations including core--core and core--valence correlation effects, i.e., cc-pwCVTZ and its variants.
The impact of the CP correction (\cref{sec:BSSE}) is also marginal for all the investigated basis sets except those without augmenting diffuse functions and without tuning for the F12 method, i.e., cc-pVTZ and cc-pwCVTZ.
Aside from these two basis sets, all the values are similar to those with heavy-aug-cc-pVTZ, employed for the main results in the present study.
The use of the heavy-aug-cc-pVTZ basis set is thus justified, as the accuracy that can be achieved by the present MLIP formalism is \qtyrange{0.1}{1}{meV/atom} and the differences from the other basis sets are hence negligible for the MLIP training.
It is also worth noting that, regarding vdW interaction, cc-pV\textit{X}Z-F12 are not optimized for vdW interaction and indeed reported to be less accurate than aug-cc-pV\textit{X}Z likely because the former lack higher-angular-momentum diffuse functions~\cite{Patkowski_JCP_137_034103_2012,Platts_JCTC_9_330_2012,Sirianni_JCTC_13_86_2017}, which also supports the use of the standard basis set of Dunning in the present study.

 \section{DFT and RPA}
\label{sec:RPA}

\begin{figure*}[tb]
\centering
\includegraphics{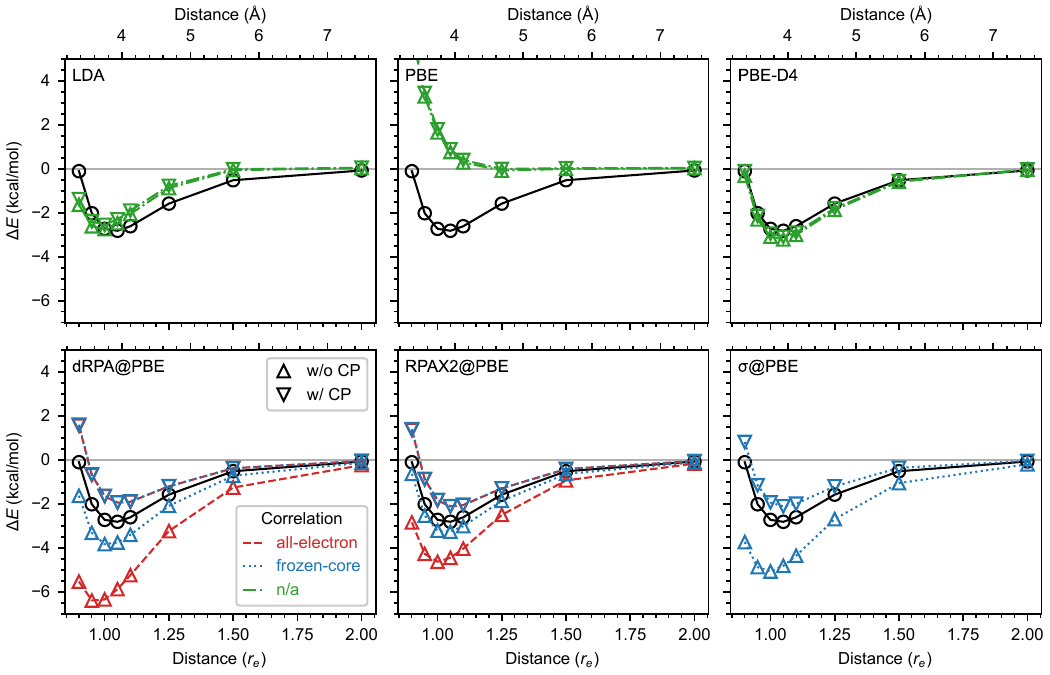}
\caption{Intermolecular interaction energies of a benzene--benzene dimer with $\pi$--$\pi$ stacking obtained with various DFT functionals and RPA-based methods (kcal/mol).
The reference CCSD(F12*)(T)/CBS values~\cite{Brauer_PCCP_2016_S66x8} are also shown for comparison in the solid black curves.
Note that all-electron treatment is not available in the $\sigma$-functional method likely because the parameters of the $\sigma$ functionals are optimized only for the frozen-core case.}
\label{fig:DFT_RPA}
\end{figure*}

\begin{table}
\centering
\caption{Intermolecular interaction energies of a benzene--benzene dimer with $\pi$--$\pi$ stacking obtained with various DFT functionals and RPA-based methods (kcal/mol).}
\label{tab:DFT_RPA}
\begin{tabular}{lcc*{8}{S[table-format=4.4,retain-explicit-plus]}}
\toprule
& & & \multicolumn{8}{c}{$r_e$} \\
\cmidrule{4-11}
\multicolumn{1}{c}{Method} &
\multicolumn{1}{c}{Correlation} &
\multicolumn{1}{c}{CP} &
\multicolumn{1}{c}{0.90} &
\multicolumn{1}{c}{0.95} &
\multicolumn{1}{c}{1.00} &
\multicolumn{1}{c}{1.05} &
\multicolumn{1}{c}{1.10} &
\multicolumn{1}{c}{1.25} &
\multicolumn{1}{c}{1.50} &
\multicolumn{1}{c}{2.00} \\
\midrule
             LDA &          n/a & w/o CP & -1.629 & -2.607 & -2.740 & -2.467 & -2.053 & -0.872 & -0.058 & +0.038 \\
             LDA &          n/a &  w/ CP & -1.371 & -2.390 & -2.543 & -2.293 & -1.898 & -0.769 & -0.015 & +0.050 \\
             PBE &          n/a & w/o CP & +6.207 & +3.300 & +1.660 & +0.768 & +0.298 & -0.075 & +0.000 & +0.033 \\
             PBE &          n/a &  w/ CP & +6.406 & +3.452 & +1.796 & +0.885 & +0.404 & -0.003 & +0.033 & +0.042 \\
          PBE-D4 &          n/a & w/o CP & -0.302 & -2.293 & -3.081 & -3.206 & -3.002 & -1.860 & -0.585 & -0.036 \\
          PBE-D4 &          n/a &  w/ CP & -0.104 & -2.141 & -2.945 & -3.088 & -2.896 & -1.788 & -0.552 & -0.027 \\
        dRPA@PBE & all-electron & w/o CP & -5.532 & -6.390 & -6.343 & -5.876 & -5.241 & -3.233 & -1.257 & -0.240 \\
        dRPA@PBE & all-electron &  w/ CP & +1.553 & -0.688 & -1.651 & -1.944 & -1.900 & -1.202 & -0.379 & -0.046 \\
        dRPA@PBE &  frozen-core & w/o CP & -1.633 & -3.312 & -3.828 & -3.752 & -3.408 & -2.086 & -0.719 & -0.130 \\
        dRPA@PBE &  frozen-core &  w/ CP & +1.601 & -0.658 & -1.632 & -1.930 & -1.889 & -1.197 & -0.376 & -0.046 \\
       RPAX2@PBE & all-electron & w/o CP & -2.838 & -4.271 & -4.630 & -4.452 & -4.033 & -2.499 & -0.931 & -0.168 \\
       RPAX2@PBE & all-electron &  w/ CP & +1.354 & -0.887 & -1.839 & -2.113 & -2.049 & -1.295 & -0.414 & -0.051 \\
       RPAX2@PBE &  frozen-core & w/o CP & -0.634 & -2.534 & -3.214 & -3.263 & -3.011 & -1.863 & -0.637 & -0.107 \\
       RPAX2@PBE &  frozen-core &  w/ CP & +1.396 & -0.861 & -1.822 & -2.102 & -2.041 & -1.291 & -0.413 & -0.051 \\
       RIRPA@PBE &  frozen-core & w/o CP & -3.734 & -4.872 & -5.071 & -4.814 & -4.352 & -2.691 & -1.054 & -0.200 \\
       RIRPA@PBE &  frozen-core &  w/ CP & +0.814 & -1.151 & -1.934 & -2.110 & -1.993 & -1.200 & -0.360 & -0.041 \\
\midrule
CCSD(F12*)T/CBS\footnotemark[1]
                 &  frozen-core &  w/ CP & -0.105 & -2.016 & -2.725 & -2.813 & -2.607 & -1.578 & -0.515 & -0.072 \\
\bottomrule
\end{tabular}
 \footnotetext[1]{Brauer~\textit{et al.}~\cite{Brauer_PCCP_2016_S66x8}}
\end{table}

The RPA was often referred to for validating the accuracy of vdW functionals like optB86b-vdW~\cite{Klimes_PRB_2011_Van}, rev-vdW-DF2~\cite{Hamada_PRB_2014_van}, SCAN+rVV10~\cite{Peng_PRX_2016_Versatile}, and \textit{r}\textsuperscript{2}SCAN+rVV10~\cite{Ning_PRB_2022_Workhorse}.
We therefore discuss the accuracy of various RPA-based methods available in MOLPRO in this appendix.
Specifically, the direct RPA~(dRPA)~\cite{Furche_PRB_2001_Molecular} is the most standard one with the exchange--correlation kernel set to zero.
The RPAX2 method~\cite{Hesselmann_PRA_2012_Random} is an extension of the RPA where exchange effects are considered in higher-order particle-hole interactions.
The $\sigma$-functional method~\cite{Trushin_JCP_154_014104_2021} is based on the adiabatic-connection-fluctuation-dissipation theorem (ACFDT).
This method is technically similar to the dRPA and employs a function of Kohn--Sham eigenvalues parametrized based on reference data obtained in experiments or in higher-accuracy computational methods.
Note that all-electron treatment seems not available in the $\sigma$-functional method with the MOLPRO~2024.1 program.

\cref{fig:DFT_RPA} and \cref{tab:DFT_RPA} present the inter-molecular interaction energy of a benzene-benzene dimer with $\pi$--$\pi$ stacking computed with the RPA-based methods on top of the orbitals obtained using the PBE functional~\cite{Perdew_PRL_77_3865_1996,*Perdew_PRL_78_1396_1997}.
The results with various DFT functionals are also shown for comparison.
The heavy-aug-cc-pVTZ basis set is employed.

Within DFT, the PBE functional fails to capture the attraction between the benzene molecules even qualitatively, in agreement with a previous study~\cite{Barone_JCC_30_934_2008}.
This well-known deficiency arises from the absence of the vdW interaction, as reported also for layered materials such as graphite~\cite{Mounet_PRB_2005_First,Barone_JCC_30_934_2008,Park_JKPS_59_196_2011,Rego_JPCM_2015_Comparative}.
The LDA appears to reproduce the reference CCSD(F12*)(T)/CBS binding curve semi-quantitatively; however, this apparent agreement is merely fortuitous, stemming from a cancellation of errors between its unphysically long-range exchange interaction and its absence of the vdW interaction~\cite{Harris_PRB_1985_Simplified,Bjoerkman_JPCM_2012_Are}.
Consequently, within the DFT framework, a semi-empirical dispersion correction such as the D4 scheme~\cite{Caldeweyher_JCP_2017_Extension, Caldeweyher_JCP_2019_generally, Caldeweyher_PCCP_2020_Extension} is necessary.

For the RPA methods, without the CP correction, the results largely depend on the methods and core-electron treatment.
For example, in the dRPA without the CP correction, the interaction energy at $r_e = 1.00$ is \qty{-3.828}{kcal/mol} in the frozen-core approximation but \qty{-6.343}{kcal/mol} in all-electron treatment, largely overestimating the interaction energy of the reference CCSD(T)/CBS value in magnitude.
With the CP correction, the interaction energies are almost independent of the core-electron treatment;
we find the interaction energies of \qtylist{-1.632;-1.651}{kcal/mol} in the frozen-core approximation and in all-electron treatment, respectively, which underestimate the reference value by \qty{1.1}{kcal/mol} in magnitude.
The RPAX2 method shows qualitatively the same trend, while the values without the CP correction and with core-electron excitation are much closer to the reference value than dRPA.
The insignificance of all-electron treatment in case with the CP correction implies that its apparent impact is indeed an artifact due to the BSSE and does not actually contribute to the London dispersion interaction.
At the same time, the large impact of the CP correction implies that this correction is essential to obtain converged results for the RPA-based methods.
Conversely, DFT functionals exhibit negligible impact of the CP correction, implying that the basis-set superposition does not substantially contribute to the description of electron density.

In previous studies, the RPA in conjunction with plane-wave basis sets showed good agreements with experiments for the inter-layer binding energy of graphite~\cite{Lebegue_PRL_105_196401_2010}, defect formation energies in silicon~\cite{Kaltak_PRB_90_054115_2014}, the melting temperatures of gold~\cite{Grabowski_PRB_91_201103_2015} and silicon~\cite{Dorner_PRL_121_195701_2018}, a phase-transition temperature of silica~\cite{Forslund_Misc_2025_Free}, etc., supporting the predictive power of the RPA for covalent, ionic, metallic, and vdW bonding in high accuracy.
With local basis sets, however, the CP correction becomes essential for the RPA, as demonstrated above, making it cumbersome to obtain basis-set converged results.
Given also that coupled‑cluster theory includes a much broader range of correlation effects than the RPA methods, PNO‑LCCSD(T)‑F12 is the method of choice for the present study.

 \clearpage
\section{Training procedure}
\label{sec:training_procedure}

\begin{figure}[hb]
\centering
\includegraphics{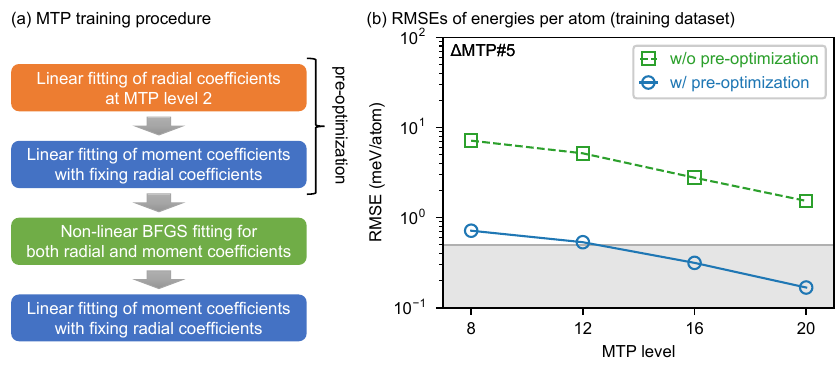}
\caption{(a) Training procedures of $\Delta$MTPs. (b) RMSEs for the training dataset \#5 without and with the pre-optimization process.}
\label{fig:training_vs_random}
\end{figure}

An MTP has essentially three types of parameters:
\begin{enumerate}
\item \texttt{radial\_coeffs}: Coefficients for radial functions denoted as $c_{\mu,z_i,z_j}^{(\beta)}$ in Eq.~(4) in Podryabinkin \textit{et al.}~\cite{Podryabinkin_JCP_159_084112_2023}.
\item \texttt{moment\_coeffs}: Coefficients for MTP basis functions denoted as $\xi_\alpha$ in Eq. (3) in the main text or in Eq.~(2) in Ref.~\cite{Podryabinkin_JCP_159_084112_2023}. 
\item \texttt{species\_coeffs}: Reference free-atom energies denoted explicitly as $V_0(z_i)$ in Eq.~(3) in the main text.
\end{enumerate}
Among them, \texttt{species\_coeffs} were fixed to the PNO-CCSD(T)-F12 values in the present study, and therefore the remaining two types of coefficients were optimized.
\cref{fig:training_vs_random}(a) shows the training procedure of the $\Delta$MTPs.
Notably, before the main non-linear BFGS fitting, the procedure involves ``pre-optimization'' of \texttt{radial\_coeffs} and \texttt{moment\_coeffs}.
Specifically, at the first step, \texttt{radial\_coeffs} were computed at MTP level 2 in a deterministic manner with linear fitting.
The linear fitting for \texttt{radial\_coeffs} is possible at MTP level 2, because the only MTP basis function is an identity scalar moment tensor, and therefore the MTP reduces to a pair-interaction potential represented by linear combinations of polynomials.
(Note that at higher MTP levels the energy depends on \texttt{radial\_coeffs} non-linearly, and hence the linear fitting is not possible.)
At the second step of pre-optimization\texttt{moment\_coeffs} were computed deterministically with linear fitting with fixing the already-initialized \texttt{radial\_coeffs}.

The performance of the pre-optimization process is demonstrated with the RMSEs for the training dataset \#5 as given in \cref{fig:training_vs_random}(b), with comparison to the results of the $\Delta$MTPs without the pre-optimization process.
The $\Delta$MTPs without the pre-optimization process shows one order larger RMSEs than those with the pre-optimization at all the investigated MTP levels.
This indicates that the pre-optimization process sets a reasonable initial guess for \texttt{radial\_coeffs} and \texttt{moment\_coeffs}, while a much more BFGS steps would be necessary to achieve the same order of RMSEs without the pre-optimization process.
This demonstrates that the pre-optimization of the radial and the moment coefficients accelerate the training of MTPs, which is therefore adopted in the present study.

 \clearpage
\section{Vibrational frequencies}
\label{sec:vibrations_full}

\cref{tab:frequencies_H2,tab:frequencies_C6H6} show the harmonic vibrational frequencies of \ce{H2} and \ce{C6H6}, respectively, computed with GFN2-xTB, various quantum-chemical methods, ANI-1ccx, and TB+$\Delta$MTP\#5.

\begin{table}[hb]
\centering
\caption{Harmonic vibrational frequencies of \ce{H2} (\unit{cm^{-1}}).\footnote[1]{The irreducible representation follows, e.g., Herzberg~\cite{Herzberg_Book_1946_Infrared} and Dresselhaus~\textit{et al.}~\cite{Dresselhaus_Book_2007_Group}}\textsuperscript{,}\footnote[2]{The heavy-aug-cc-pVTZ basis set was used for quantum-chemical calculations.}}
\label{tab:frequencies_H2}
\begin{tabular}{l*{1}{S[table-format=5.3]}}
\toprule
\multicolumn{1}{c}{Method} &
$\Sigma_\mathrm{g}^{+}$ \\
\midrule
GFN2-xTB               & 3755 \\
\midrule
HF                     & 4589 \\
\midrule
LDA                    & 4184 \\
PBE                    & 4317 \\
PBE-D4                 & 4317 \\
\midrule
MP2                    & 4528 \\
MP2-F12                & 4521 \\
CCSD                   & 4411 \\
CCSD-F12               & 4414 \\
\midrule
ANI-1ccx               & 2867 \\
\bfTBDMTP[5]           & 4411 \\
\midrule
Exp.\footnotemark[3]   & 4401.213 \\
\bottomrule
\end{tabular}
 \footnotetext[3]{Harmonic vibrational frequencies in Huber and Herzberg~\cite{Huber_Book_1979_Molecular}.}
\end{table}

\begin{table*}[hbp]
\centering
\caption{Harmonic vibrational frequencies of \ce{C6H6} (\unit{cm^{-1}}).\footnote[1]{The mode numbers follow Wilson~\cite{Wilson_PR_45_706_1934}.}\textsuperscript{,}\footnote[2]{The irreducible representations follow, e.g., Mulliken~\cite{Mulliken_PR_1933_Electronic,Mulliken_JCP_23_1997_1955,*Mulliken_JCP_24_1118_1956},
Tisza~\cite{Tisza_ZP_82_48_1933},
Herzberg~\cite{Herzberg_Book_1946_Infrared},
Dresselhaus~\textit{et al.}~\cite{Dresselhaus_Book_2007_Group}, and Bradley and Cracknell~\cite{Bradley_Book_2010_Mathematical}.}\textsuperscript{,}\footnote[3]{The heavy-aug-cc-pVTZ basis set was used for quantum-chemical calculations.}}
\label{tab:frequencies_C6H6}
\begin{tabular}{lc*{10}{S[table-format=5.2]}}
\toprule
&
&
\multicolumn{1}{c}{ 1} &
\multicolumn{1}{c}{ 2} &
\multicolumn{1}{c}{ 3} &
\multicolumn{1}{c}{ 4} &
\multicolumn{1}{c}{ 5} &
\multicolumn{1}{c}{ 6} &
\multicolumn{1}{c}{ 7} &
\multicolumn{1}{c}{ 8} &
\multicolumn{1}{c}{ 9} &
\multicolumn{1}{c}{10} \\
\cmidrule(){3-12}
\multicolumn{1}{c}{Method} &
\multicolumn{1}{c}{Correlation} &
\multicolumn{1}{c}{A\textsubscript{1g}} &
\multicolumn{1}{c}{A\textsubscript{1g}} &
\multicolumn{1}{c}{A\textsubscript{2g}} &
\multicolumn{1}{c}{B\textsubscript{2g}} &
\multicolumn{1}{c}{B\textsubscript{2g}} &
\multicolumn{1}{c}{E\textsubscript{2g}} &
\multicolumn{1}{c}{E\textsubscript{2g}} &
\multicolumn{1}{c}{E\textsubscript{2g}} &
\multicolumn{1}{c}{E\textsubscript{2g}} &
\multicolumn{1}{c}{E\textsubscript{1g}} \\
\midrule
GFN2-xTB               & n/a          & 1067   & 3093   & 1304   &  658   &  937   & 579   & 3072   & 1600   & 1198   &  882   \\
\midrule
HF                     & n/a          & 1071   & 3349   & 1498   &  768   & 1129   & 662   & 3320   & 1770   & 1281   &  957   \\
\midrule
LDA                    & n/a          & 1017   & 3110   & 1317   &  705   &  985   & 602   & 3088   & 1619   & 1150   &  833   \\
PBE                    & n/a          &  995   & 3123   & 1332   &  703   &  985   & 599   & 3097   & 1590   & 1159   &  837   \\
PBE-D4                 & n/a          &  995   & 3123   & 1332   &  703   &  985   & 599   & 3097   & 1590   & 1159   &  837   \\
\midrule
MP2                    & frozen-core  & 1011   & 3232   & 1369   &  655   &  984   & 606   & 3205   & 1633   & 1192   & 861 \\
                       & all-electron & 1028   & 3234   & 1373   &  700   & 1073   & 607   & 3193   & 1640   & 1188   & 885 \\
MP2-F12                & frozen-core  & 1014   & 3233   & 1377   &  705   & 1018   & 610   & 3210   & 1640   & 1196   & 867 \\
                       & all-electron & 1016   & 3228   & 1380   &  728   & 1044   & 612   & 3205   & 1645   & 1195   & 867 \\
CCSD                   & frozen-core  & 1023   & 3229   & 1390   &  666   & 1008   & 617   & 3200   & 1666   & 1203   & 876 \\
                       & all-electron & 1040   & 3232   & 1395   &  708   & 1106   & 618   & 3186   & 1672   & 1199   & 902 \\
CCSD-F12               & frozen-core  & 1026   & 3231   & 1397   &  712   & 1034   & 620   & 3205   & 1672   & 1206   & 883 \\
                       & all-electron & 1029   & 3225   & 1398   &  723   & 1063   & 621   & 3195   & 1676   & 1204   & 885 \\
CCSD(T)                & frozen-core  & 1002   & 3202   & 1369   &  646   &  976   & 606   & 3173   & 1630   & 1186   & 855 \\
                       & all-electron & 1020   & 3205   & 1374   &  690   & 1074   & 608   & 3160   & 1637   & 1182   & 881 \\
CCSD(T)-F12            & frozen-core  & 1006   & 3205   & 1377   &  692   & 1003   & 610   & 3179   & 1638   & 1189   & 862 \\
                       & all-electron & 1009   & 3199   & 1377   &  705   & 1033   & 610   & 3169   & 1641   & 1187   & 864 \\
\midrule
ANI-1ccx               & n/a          & 1062   & 3272   & 1367   &  616   & 1014   & 621   & 3230   & 1693   & 1210   &  882   \\
\bfTBDMTP[5]           & n/a          & 1010   & 3219   & 1395   &  716   & 1004   & 606   & 3172   & 1629   & 1188   &  887   \\
\midrule
Exp.\footnotemark[4]   & n/a          & 1008   & 3208   & 1390   &  718   & 1011   & 613   & 3191   & 1639   & 1192   &  866   \\
\midrule
&
&
\multicolumn{1}{c}{11} &
\multicolumn{1}{c}{12} &
\multicolumn{1}{c}{13} &
\multicolumn{1}{c}{14} &
\multicolumn{1}{c}{15} &
\multicolumn{1}{c}{16} &
\multicolumn{1}{c}{17} &
\multicolumn{1}{c}{18} &
\multicolumn{1}{c}{19} &
\multicolumn{1}{c}{20} \\
\cmidrule(){3-12}
\multicolumn{1}{c}{Method} &
\multicolumn{1}{c}{Correlation} &
\multicolumn{1}{c}{A\textsubscript{2u}} &
\multicolumn{1}{c}{B\textsubscript{1u}} &
\multicolumn{1}{c}{B\textsubscript{1u}} &
\multicolumn{1}{c}{B\textsubscript{2u}} &
\multicolumn{1}{c}{B\textsubscript{2u}} &
\multicolumn{1}{c}{E\textsubscript{2u}} &
\multicolumn{1}{c}{E\textsubscript{2u}} &
\multicolumn{1}{c}{E\textsubscript{1u}} &
\multicolumn{1}{c}{E\textsubscript{1u}} &
\multicolumn{1}{c}{E\textsubscript{1u}} \\
\midrule
GFN2-xTB               & n/a          &  693   &  957   & 3069   & 1320   & 1176   & 368   &  931   & 1090   & 1460   & 3084   \\
\midrule
HF                     & n/a          &  760   & 1096   & 3308   & 1339   & 1175   & 451   & 1108   & 1128   & 1633   & 3338   \\
\midrule
LDA                    & n/a          &  658   &  996   & 3078   & 1391   & 1123   & 396   &  957   & 1041   & 1467   & 3102   \\
PBE                    & n/a          &  665   &  991   & 3087   & 1343   & 1136   & 396   &  957   & 1034   & 1465   & 3113   \\
PBE-D4                 & n/a          &  665   &  991   & 3087   & 1343   & 1136   & 396   &  957   & 1034   & 1465   & 3113   \\
\midrule
MP2                    & frozen-core  &  688   & 1015   & 3191   & 1463   & 1165   & 399   &  983   & 1058   & 1503   & 3222 \\
                       & all-electron &  708   & 1030   & 3168   & 1486   & 1162   & 408   & 1033   & 1063   & 1513   & 3221 \\
MP2-F12                & frozen-core  &  690   & 1026   & 3200   & 1455   & 1167   & 409   &  988   & 1063   & 1509   & 3226 \\
                       & all-electron &  691   & 1034   & 3195   & 1459   & 1167   & 411   &  996   & 1064   & 1512   & 3221 \\
CCSD                   & frozen-core  &  699   & 1024   & 3187   & 1313   & 1161   & 406   & 1005   & 1069   & 1528   & 3218 \\
                       & all-electron &  720   & 1040   & 3162   & 1342   & 1159   & 416   & 1059   & 1074   & 1539   & 3217 \\
CCSD-F12               & frozen-core  &  700   & 1036   & 3194   & 1317   & 1163   & 414   & 1014   & 1073   & 1534   & 3221 \\
                       & all-electron &  702   & 1043   & 3182   & 1326   & 1162   & 414   & 1025   & 1074   & 1536   & 3214 \\
CCSD(T)                & frozen-core  &  683   & 1009   & 3160   & 1333   & 1155   & 395   &  978   & 1051   & 1503   & 3191 \\
                       & all-electron &  705   & 1024   & 3134   & 1361   & 1152   & 405   & 1033   & 1056   & 1513   & 3190 \\
CCSD(T)-F12            & frozen-core  &  685   & 1021   & 3168   & 1337   & 1157   & 403   &  987   & 1056   & 1509   & 3196 \\
                       & all-electron &  687   & 1028   & 3156   & 1346   & 1156   & 403   &  999   & 1056   & 1511   & 3188 \\
\midrule
ANI-1ccx               & n/a          &  789   &  998   & 3220   & 1290   & 1162   & 390   &  994   & 1074   & 1487   & 3245   \\
\bfTBDMTP[5]           & n/a          &  673   & 1017   & 3177   & 1296   & 1175   & 407   &  975   & 1066   & 1508   & 3186   \\
\midrule
Exp.\footnotemark[4]   & n/a          &  686   & 1024   & 3172   & 1318   & 1167   & 407   &  989   & 1058   & 1512   & 3191   \\
\bottomrule
\end{tabular} \footnotetext[4]{Harmonic vibrational frequencies estimated by Handy~\textit{et al.}~\cite{Handy_CPL_197_506_1992} based on the fundamental vibrational frequencies of Goodman~\textit{et al.}~\cite{Goodman_JPC_95_9044_1991}}
\end{table*} \clearpage
\section{DFT for the \texorpdfstring{\ce{C48H30}}{C48H30} COF}
\label{sec:VASP}

For comparison with GFN2-xTB and TB+$\Delta$MTP for the \ce{C48H30} COF in \textcolor{violet}{Sec.~III\,F} in the main text, the calculations with a vdW DFT functional, specifically the PBE functional with
the D4 correction~\cite{Caldeweyher_JCP_2019_generally}, were also conducted using the VASP code~\cite{Kresse_JNS_1995_Ab,Kresse_CMS_1996_Efficiency,Kresse_PRB_1999_ultrasoft}.
Plane-wave basis sets were employed in conjunction with the plane-wave projector augmented wave (PAW) method~\cite{Bloechl_PRB_1994_Projector}.
The plane-wave cutoff energy was set to \SI{520}{\electronvolt}.
Reciprocal spaces of the six-layer supercell models, including 468 atoms, were sampled by a $\Gamma$-point-only mesh in conjunction with the Gaussian smearing with a width of \qty{0.05}{eV}.
The 2s2p orbitals of C and the 1s orbital of H were treated as valence states.
Total energies were minimized until they converged within \qty{1e-7}{\electronvolt} per simulation cell for each self-consistent-field cycle.
Structure relaxations were performed until all forces on atoms converged within \SI{5e-3}{\electronvolt/\angstrom}.

\begin{figure}[hbt]
\centering
\includegraphics{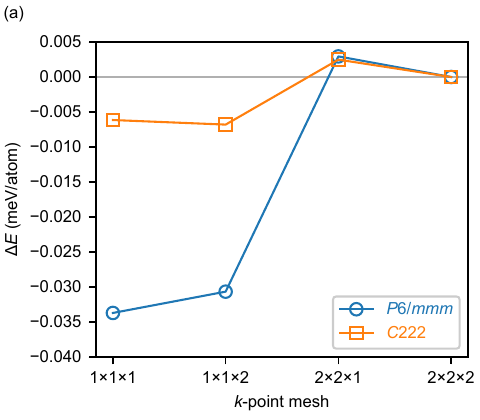}
\includegraphics{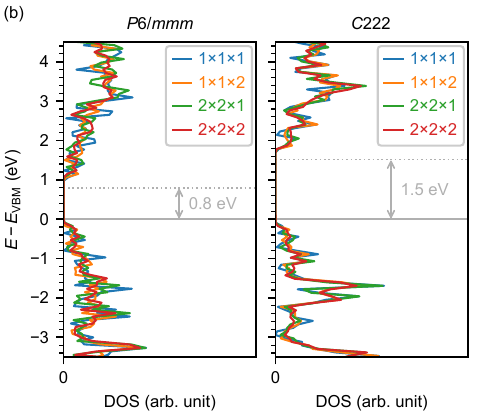}
\caption{(a) Energies of the 468-atom supercell models of the \ce{C48H30} COF with different \textit{k}-point-mesh densities.
For each structure, the energy with the \numproduct{2x2x2} \textit{k}-point mesh is set as the reference.
(b) Electronic DOSs of the \ce{C48H30} COF with different $k$-point-mesh densities.}
\label{fig:k_convergence}
\end{figure}

The convergences of the energies and the electronic DOSs with respect to $k$-point-mesh density are tested for the structures optimized with the $\Gamma$-point-only meshes.
Specifically,
\numproduct{1x1x1},
\numproduct{1x1x2},
\numproduct{2x2x1}, and
\numproduct{2x2x2} meshes were considered, where the first two and the last one numbers correspond to the in-plane and the out-of-plane directions, respectively.
The results are summarized in \cref{fig:k_convergence}.

The energy differences between the $\Gamma$-point-only meshes and the \numproduct{2x2x2} meshes are less than \qty{0.04}{meV/atom} for the $P6/mmm$ structure and \qty{0.01}{meV/atom} for the $C222$ structure, which are one-order smaller than the fitting error of $\Delta$MTP\#5 (\qty{0.4}{meV/atom}, cf.~\textcolor{violet}{Sec. III\,A} in the main text) and three-order smaller than the energy difference between the $P6/mmm$ and the $C222$ structures (\qty{10.7}{meV/atom}, cf.~\textcolor{violet}{Sec.~III\,F} in the main text).

The electronic DOSs with the $\Gamma$-point-only meshes also reproduces the characteristic features of the DOSs with the higher $k$-point-mesh densities, such as the band gaps and the peak positions.
The band gaps in PBE-D4 are calculated to be \qtylist{0.8;1.5}{eV} for the $P6/mmm$ and the $C222$ structures, respectively, although we note that band gaps in DFT are typically underestimated.

